\pdfoutput=1

\documentclass[11pt]{article}

\usepackage[preprint]{acl}

\usepackage{times}
\usepackage{latexsym}

\usepackage[T1]{fontenc}

\usepackage[utf8]{inputenc}
\usepackage{booktabs}
\usepackage{multirow}
\usepackage{multicol}
\usepackage{microtype}

\usepackage{inconsolata}

\usepackage{graphicx}
\newcommand{\eat}[1]{}

\title{\textbf{Hindi-BEIR} : \textit{A Large Scale Retrieval Benchmark in Hindi}}

\author{
 \textbf{Arkadeep Acharya\textsuperscript{1}} \thanks{Work done as an intern at IBM Research},
 \textbf{Rudra Murthy\textsuperscript{2}},
 \textbf{Vishwajeet Kumar\textsuperscript{2}},
 \textbf{Jaydeep Sen\textsuperscript{2}},
\\
 \textsuperscript{1}Department of Computer Science and Engineering, Indian Institute of Technology Patna,\\
 \textsuperscript{2}IBM Research
\\
 arkadeep\_2101ai41@iitp.ac.in,\{rmurthyv,vishk024,jaydesen\}@in.ibm.com
}

\begin{document}
\maketitle
\begin{abstract}

Given the large number of Hindi speakers worldwide, there is a pressing need for robust and efficient information retrieval systems for Hindi. Despite ongoing research, there is a lack of comprehensive benchmark for evaluating retrieval models in Hindi. To address this gap, we introduce the Hindi version of the BEIR\footnote{\url{https://github.com/beir-cellar/beir}} benchmark, which includes a subset of English BEIR datasets translated to Hindi, existing Hindi retrieval datasets, and synthetically created datasets for retrieval. The benchmark is comprised of $15$ datasets spanning across $8$ distinct tasks. We evaluate state-of-the-art multilingual retrieval models on this benchmark to identify task and domain-specific challenges and their impact on retrieval performance. By releasing this benchmark and a set of relevant baselines, we enable researchers to understand the limitations and capabilities of current Hindi retrieval models, promoting advancements in this critical area. The datasets from Hindi-BEIR are publicly available at \href{https://huggingface.co/collections/ArkaAcharya/datasets-667004c0dc348adcabc629be}{Hindi-BEIR}

\end{abstract}

\section{Introduction} 
Retrieval models are essential for efficiently accessing and utilizing the vast amounts of information available today. These models enable quick and accurate extraction of relevant data, which is crucial for informed decision making in various downstream applications across multitudes of domains. In this modern era of LLMs, retrievers have become even more relevant in curbing hallucinations in LLM generations through RAG~\cite{lewis2021retrievalaugmented} style architectures. Research in retrievers have long been predominantly focused on English language with availability of necessary comprehensive benchmarks such as BEIR~\cite{thakur2021beir}. Although, more recently there have been efforts to build multi-lingual retrievers and associated evaluation benchmarks ~\cite{10.1162/tacl_a_00595,zhang-etal-2021-mr} spanning non-English languages too, multi-lingual expansion of robust benchmarks is still a work in progress, with many languages lacking a BEIR like comprehensive benchmark. In this work, we take on one such language Hindi and contribute a new benchmark called \textbf{Hindi-BEIR}: \textit{A Large Scale Retrieval Benchmark in Hindi}. We will explain our reasoning behind choosing Hindi for this benchmark creation as well as provide details of the challenges and important considerations needed in creating a retrieval benchmark on a new language.

Hindi is the 3rd most spoken language in the world with over half a billion speakers worldwide \footnote{\url{https://www.thecollector.com/what-are-the-most-spoken-languages-in-the-world/}} and is one of the official languages of India, therefore, used in a large number of research and practical industry use-cases, along with other indic languages\footnote{\url{https://en.wikipedia.org/wiki/Indo-Aryan_languages}}. The progress in Indic language datasets so far has been limited to single domain single task datasets ~\cite{haq2023indicirsuite}, but more focused towards including multiple of Indic languages. However, because of their lack of domain and task diversity, these datasets are not at all capable of evaluating the current state of retriever models on practical applications with varying document sizes and contents. 

In this work, we take the position that creating a BEIR like diversified and robust benchmark is the necessary first step to track and advance any retrieval research effectively in a new language and such a benchmark will be more useful than ad-hoc datasets with multiple languages. We choose the most widely spoken language among Indic languages Hindi as our pivot language to pioneer a large scale diversified retrieval benchmark Hindi-BEIR, spanning 15 diverse datasets across \textbf{$8$} tasks and over $5$ distinct domains. \textbf{Hindi-BEIR} aims to achieve two key objectives: (1) Establish a standardized retrieval benchmark to assess, compare and advance the state-of-the-art retriever models and (2) Provide workable insights into the potential research directions on retrieval models in Hindi.\par

Besides being a language of high usage, a retrieval benchmark in Hindi offers very unique challenges too such as :\\
    $\bullet$\textbf{ Script Difference:} Hindi uses the Devanagari script, which is fundamentally different from the Latin script used in English. This affects character encoding, text normalization, and processing. Existing tokenizers trained on English data may not handle Hindi text well, thus needing to evaluate various tokenization strategies too.
    
    \eat{Tokenizing Hindi text poses unique challenges due to its relatively rich morphology. Existing tokenizers trained on English data may not handle Hindi text effectively, leading to suboptimal retrieval performance. }
    
    

    $\bullet$\textbf{ Grammatical Structure:} Hindi grammar follows different syntactical rules e.g. using Subject-Object-Verb (SOV) order, unlike the Subject-Verb-Object (SVO) structure in English. Hindi words often include more inflections and agglutinations, affecting word tokenization and testing the robustness of retrieval models. 

    $\bullet$ \textbf{ Ambiguity:} In Hindi, some proper nouns can also function as common nouns. For instance, the name \textit{Lata}, a common female name, can also refer to \textit{creeper}, a common noun. A query like \textit{lata ko kaise saaph karen} (how to clean a creeper) can easily mislead lexicon based retrieval systems and test a model's ablity for word sense disambiguation. 
    

We finally sum up our contributions as follows:
\begin{itemize}
\item We introduce \textbf{Hindi-BEIR}, the first comprehensive retrieval benchmark in the Hindi language, encompassing 15 diverse datasets from 6 distinct domains. This was accomplished through the harmonization of high-quality translated data from BEIR, the conversion of existing Hindi datasets for retrieval purposes, and the introduction of strategically generated synthetic data.
 
\item We take current day most popular multilingual retriever models and compare their baseline performance on \textbf{Hindi-BEIR} to highlight important insights and scope of future research.
\end{itemize}

 \begin{table}[!htb]
    \centering
    \resizebox{0.5\textwidth}{!}{%
    \begin{tabular}{llrr}
    \toprule
      \textbf{Dataset Name} & \textbf{Tasks}   & \textbf{\#Corpus} & \textbf{\# Queries}  \\
    \midrule
       ArguAna & Argument Retrieval & 7763 & 1194 \\
       FiQA-2018 & Question-Answering &48178 & 5924 \\
       TREC-COVID & Bio-Medical IR &76492 & 49 \\
       SCIDOCS & Citation-Prediction & 22050 & 850  \\
       SciFact & Fact-Checking & 2849 & 1099 \\
       Touch\'{e}-2020 & Argument Retrieval & 355273 & 49 \\
        NQ & Question Answering & 2595865  &2952  \\
        FEVER & Fact Checking & 5362876 & 120075\\
        Climate-FEVER & Fact Checking & 5362911 & 1499 \\
        \midrule
        CC News Retrieval & News Article Retrieval& 5005483 & 49699  \\
        Bhaav Retrieval & Story Comprehension (Q/A) & 1536 & 34  \\
        \midrule
        MLDR &Long Document Retrieval& 3806 & 200  \\
        MIRACL & Question Answering & 3494 & 350 \\
        IndicQARetrieval &Question Answering & 1547 & 1547 \\
        mMARCO &Passage Retrieval& 8841823 & 6980  \\
       WikiPediaRetrieval & Question Answering & 13500 & 1500 \\
    \bottomrule
    \end{tabular}%
    }
    \caption{Statistics of the Dataset in the Hindi-BEIR Benchmark showing the tasks which each dataset covers and the number of corpus and  query in the evaluation set of each dataset in the Hindi-BEIR Benchmark.}
    \label{tab:dataset_stats}
\end{table}

\section{Related Work}
\eat{There have been past works in building non-English retriever benchmarks dedicated to specific languages such as ~\cite{louis2022statutory} or in general multi-lingual benchmarks such as ~\cite{zhang2022making,chen2024bge}. Also, most of the popular Indic language datasets ~\cite{kumar-etal-2022-indicnlg,sabane2024breaking} out in public are for generation or broadly QA tasks. To be relevant to the current work, we review retrieval benchmarks specifically built around Indic languages. 
}

Being relevant to the current work, we review only the retrieval benchmarks for Indic languages. One of the known source of testing retriever performance on Indic languages was to test on the Indic language subset contained in  multi-lingual IR datasets such as MIRACL~\cite{zhang2022making}, MR.Tydi~\cite{zhang-etal-2021-mr} having instances from Indic languages such as Hindi, Bengali, and Telugu. \eat{Note that these benchmarks weren't exclusively prepared for Indic languages but in their effort for multi-lingual reach some of the more popular Indic languages got covered. For example, MIRACL, Mr. Tydi contained instances from Indic languages such as Hindi, Bengali, and Telugu.} However, two major drawbacks here were: (1) all of these subsets were specific to a domain, hence couldn't really be used as a robust benchmark and (2) the indic language specific subsets of instances didn't contain enough data points to facilitate building neural retrieval models for indic languages either. 

Forum for Information Retrieval Evaluation (FIRE) \cite{DBLP:conf/fire/2011,DBLP:conf/fire/2013,DBLP:conf/fire/2014,DBLP:conf/fire/2015,DBLP:conf/fire/2016,DBLP:conf/fire/2017,DBLP:conf/fire/2018,DBLP:conf/fire/2019,DBLP:conf/fire/2020,DBLP:conf/fire/2021,DBLP:conf/fire/2022,DBLP:conf/fire/2023} has been actively conducting shared-tasks, inviting papers centered around information retrieval in Indian languages. The datasets used as part of the shared tasks are from newspaper articles. More importantly, the datasets from the FIRE evaluation tracks are not publicly available, which may limit their usability.

More recently a relatively large scale retrieval benchmark MSMarco~\cite{DBLP:journals/corr/NguyenRSGTMD16} got translated into multiple languages to produce mMARCO~\cite{bonifacio2022mmarco} , including Hindi from the set of Indic languages. mMARCO was further extended to more Indic languages by IndicIRSuite~\cite{haq2023indicirsuite} which is the most recent retrieval benchmark in Indic languages. However, the common problem for both mMARCO and  IndicMarco is that these benchmarks too are tied to a single type of retrieval task and web domain as was originally in MSMarco and therefore, do not provide a robust benchmark. To the best of our knowledge, Hindi-BEIR represents the first comprehensive benchmark which spans diverse domains and tasks and therefore, provides the first BEIR~\cite{thakur2021beir} equivalent comprehensive retrieval benchmark in Hindi.

\section{Hindi-BEIR Dataset} \label{dataset}

Table \ref{tab:dataset_stats} summarises the various datasets included in Hindi-BEIR benchmark along with the number of documents, queries, and, domain information.

To ensure that the Hindi-BEIR Benchmark is comprehensive, challenging, and accessible to the public for future research and evaluation, we adhered to the following objectives:\textbf{1) Diverse Domains and Tasks:} We include datasets from diverse domains like Wikipedia and news articles to niche domains like scientific publications to test the robustness and generalization ability of the retrieval models. Additionally, a good retrieval model should handle documents and queries of varying lengths equally well. We have ensured that the datasets included in Hindi-BEIR benchmark exhibit a wide range of query and document lengths among the datasets. \textbf{2) Difficulty Level:} We ensure that the datasets are challenging and systems relying solely on lexical overlap have a hard time retrieving the correct document.\textbf{3) Public Availability: } All datasets curated in Hindi-BEIR Benchmark will be made publicly available. 

We discuss the Hindi-BEIR benchmark creation in the following subsections.
Please refer to Appendix \ref{subsec:dataset_desc} for more detailed discussion.
\subsection{Translating English Datasets to Hindi:}
\label{subsec:translation}
We translate a subset of the existing English datasets from the BEIR benchmark into Hindi. This approach was necessitated by the need to ensure comprehensive coverage across multiple domains and to maintain a high level of data quality and complexity. We utilized the Indic-Trans2 model \footnote{Our choice for IndicTrans2 over other translation models have been discussed in \ref{subsec:faq} } \cite{gala2023indictrans}, a multilingual NMT model supporting translations across all 22 scheduled Indic languages (including English). We employ back-translation technique to retain good translations. Specifically, given an English query/document we translate it to Hindi. This Hindi-translated query/document is translated back to English. We calculate the Chrf(++) score \cite{popovic-2017-chrf} between the original English query/document and the backtranslated English query/document. We retained only those translations with a Chrf++ score exceeding a threshold. We empirically set the threshold to 50 after manually verifying the translation quality of texts obtained from different thresholds. 

This strategy enables us to leverage the wealth of existing high-quality datasets in English while making them accessible and useful for Hindi language information retrieval tasks. 9 out of the 15 datasets which includes Arguana, FiQA-2018, TREC-COVID, SCIDOCS, SciFact, Touch\'e-2020, NQ, FEVER, and, Climate-FEVER were created in by this method.

\subsection{Creation of Retrieval datasets from existing datasets}
\label{cc_news}
We create Cross-lingual Hindi CC News Retrieval datasets from existing datasets.

\textbf{Hindi CC News Retrieval :} This dataset is derived from the Hindi subset of the Multilingual CC News dataset\footnote{\url{https://huggingface.co/datasets/intfloat/multilingual_cc_news}}, which contains 7,444,584 data points comprising news articles and their corresponding titles. The article text serves as the corpus to be retrieved, with the titles acting as queries. 


We selected only those title-article pairs with a Rouge-L score \cite{lin-2004-rouge} of 0 between the title and article, thus reducing the dataset size to 5,005,483. The article text obtained after this filtering step constituted the total corpus of our dataset. Next, we further consider only those title-article pairs with a Jaccard index\footnote{\url{https://en.wikipedia.org/wiki/Jaccard_index}} of 0 between the title and article, thus reducing the dataset size to 451,803. This resulted in English titles with Hindi articles making the retrieval task challenging as the model needs to rely on cross-lingual understanding. From these filtered data points, we randomly selected 49,699 queries to form the test split, which has been incorporated into the Hindi-BEIR benchmark. Additionally, we released training and validation splits consisting of 301,578 and 100,526 query-corpus pairs, respectively. The train, test, and validation splits collectively encompass all data points obtained after the second filtering step.

\subsection{Compiling existing Multilingual Datasets}
We also include 5 publicly available datasets namely MIRACL \cite{zhang-etal-2021-mr}, MLDR  \cite{chen2024bge}, mMarco \cite{bonifacio2022mmarco}, IndicQARetrieval (developed by transforming the IndicQA Dataset \cite{doddapaneni2023leaving} for a retrieval task) and WikiPediaRetrieval \footnote{\url{https://huggingface.co/datasets/ellamind/wikipedia-2023-11-retrieval-multilingual-queries}}. 

\section{Experimental Setup}
\label{sec:results}
 \begin{table}[!htb]
    \centering
    \resizebox{0.5\textwidth}{!}{%
    \begin{tabular}{lrrrrr}
    \toprule
      \textbf{Dataset Name} & \textbf{BGE-M3}   & \textbf{mE5} & \textbf{LASER} & \textbf{LaBSE} &  \textbf{BM-25}\\
    \midrule
       ArguAna & \textbf{53.81} &49.96 & 11.27 & 32.90&  43.75 \\
       FiQA-2018 & \textbf{25.89} & 22.38 & 1.58 & 7.23&  16.57\\
       TREC-COVID & \textbf{64.60} &62.42 & 3.95 & 29.94&  52.30 \\
       SCIDOCS & \textbf{14.24} & 10.42 & 0.59 & 6.95 &  11.40  \\
       SciFact & \textbf{52.39} & 51.50 & 5.37 & 33.42  &  60.80\\
       Touch\'{e}-2020 & 26.68 & \textbf{27.44} & 1.06 & 6.82  & 33.59\\
        NQ & 39.15 & \textbf{44.10} & 0.49 & 9.36 &  16.79\\
        FEVER & \textbf{66.91} & 32.87 & 0.19 & 8.27 &  40.57\\
        Climate-FEVER & \textbf{23.71} & 5.93 & 0.28 & 3.72 & 14.00\\
        \midrule
        CC News Retrieval & \textbf{34.40} & 20.81 & 0.52 & 5.63 & 0.01 \\
        \midrule
        MLDR & \textbf{40.91} & 19.08 & 0.18 & 7.65 &  43.70 \\
        MIRACL & \textbf{59.34} & 58.11 & 0.69 & 13.76 &  40.98\\
        IndicQARetrieval & \textbf{69.92} & 67.11 & 21.28 & 46.85 &  74.02\\
        mMARCO & 29.49& \textbf{29.94} & 0.48 & 6.98 &  16.53\\
       WikiPediaRetrieval & \textbf{87.38} & 84.40 & 0.03 &61.28 &  82.54 \\
       \midrule
       \textbf{Average} & \textbf{47.29} & 39.93 & 3.26 &18.72 & 36.50\\
    \bottomrule    
    \end{tabular}%
    }
    \caption{NDCG@10 scores of existing multilingual model on Hindi-BEIR datasets}
    \label{tab:results}
\end{table}

We evaluate existing multilingual dense retrievers on our Hindi-BEIR benchmark. \textbf{Multilingual E5 (mE5) \cite{wang2024multilingual}} was developed by continually pre-training the E5 model \cite{wang2024text} on a large multilingual corpus using a weakly supervised contrastive pretraining method with InfoNCE contrastive loss \cite{oord2019representation}. It was then fine-tuned on high-quality labeled multilingual datasets for retrieval tasks. mE5 has a context length of 512 tokens.

The \textbf{BGE-M3 \cite{chen2024bge}} model was pre-trained on a large multilingual and cross-lingual unsupervised data, and subsequently fine-tuned on high-quality multilingual retrieval datasets using a custom loss function based on the InfoNCE loss function. BGE-M3 supports a context length of 8,192 tokens. 

\textbf{LASER\cite{Artetxe_2019}} focuses on universal language agnostic sentence embeddings across 93 different languages. LASER uses a language-agnostic BiLSTM encoder architecture trained on parallel corpora from different languages, without any retrieval-specific fine-tuning. 

\textbf{LaBSE \cite{feng2022languageagnostic}} uses a dual encoder model using BERT for obtaining language-agnostic sentence embedding, also without any retrieval-specific fine-tuning. It has a maximum context length of 256 tokens.

We extend the MTEB benchmark code repository\footnote{\url{https://github.com/embeddings-benchmark/mteb}} to include Hindi-BEIR and evaluate the models discussed above. We report NDCG@10 \cite{Jrvelin2002CumulatedGE} numbers.

\section{Results}
We outline most interesting observations from the results as follows:

\textbf{Retrieval specific finetuning helps:} Table \ref{tab:results} can be seen to have two segments. BGE-M3, mE5 which has seen retrieval specific fine-tuning vastly outperforms task agnostic sentence embedding models LASER and LaBSE. 

\eat{in terms of performance We observe that retrieval models like BGE-M3 and mE5 models outperform LASER and LABSE by a significant margin. Except for Touch\'{e}-2020 and NQ datasets, BGE-M3 model outperforms mE5 model sometimes by a significant margin. On datasets from niche domains like scientific publications (Scidocs, Climate-FEVER), financial domains (FiQA-2018) existing models exhibit poor retrieval performance.}

\textbf{Performance variance across tasks/domains:} One of the important aspects of building Hindi-BEIR was to see how a retriever adapts to different domains and tasks. As we can see, mE5 specifically struggles in fact checking datasets with average NDCG@10 of $30\%$ much below than its overall average, while BGE-M3 although not great, still fares much better for fact checking task. For citation prediction task on SciDocs, both the models perform quite poor. Also, for niche domains such as Finance and Climate, we see a sharp drop for both the models, prompting the need for focused research in these areas.

\textbf{mE5 Struggles with long document retrieval :} mE5 performs substantially poor in MLDR dataset, indicating its sub-par performance when dealing with long documents. This serves as an empirical proof that shorter context length of 512 tokens hurts mE5's performance compared to BGE-M3 which can handle 8192 tokens. 

\textbf{Vector dimension, context length matters} Except for Touch\'{e} and NQ, BGE-M3 outperforms mE5 in general with an average improvement of ~$7\%$ percentage points. Both of the models having seen sufficient amount of pre-training and fine-tuning data, this improvement is possibly because BGE-M3 uses higher dimension vectors of length 1024 and supports higher context length of 8192 as compared to ME5 with 768 dim vectors and 512 context length. 

 \textbf{Poor Performance of BM-25 on CC-News Retrieval: }  BM25 has very poor performance on CC News Retrieval datasets only. CC News Retrieval dataset, was intentionally designed to test the ability of multilingual dense retrieval models in learning language agnostic embeddings for retrieval. This dataset presents a scenario where the query is in English and the corpus is in Hindi that needs to move beyond token level matching. This explains the poor performance of BM25 on this dataset due to the lack of overlap between English and Hindi tokens.

\eat{
\textbf{Poor performance of mE5 on task involving fact checking :} Evaluations of retrieval models on fact-checking datasets like Fever and Climate-Fever underlines the poor performance. Nevertheless, between BGE-M3 and mE5 the number shows a clear advantage of BGE-M3 over mE5 on this task. 

\textbf{Mixed performance of mE5 and BGE M3 of Q/A tasks :} Table \ref{tab:results} highlights the mixed performance of BGE-M3 and mE5 models on Question-Answering (Q/A) task. mE5 claims an advantage over BGE-M3 model on the NQ dataset. The results for Fiqa-2018 and IndicQARetrieval are close with BGE-M3 showing a slight advantage.

\textbf{Poor performance on citation prediction tasks:} Both BGE-M3 and mE5 models performs poorly when it comes to citation prediction tasks on the SCIDOCS dataset. 
}

\section{Conclusion}
 We introduce Hindi-BEIR benchmark, the first comprehensive IR benchmark in Hindi language consisting of 15 datasets, over 27 Million documents as corpus and nearly 200K queries, covering 5 different domains and 8 distinct tasks. Performance analysis of current day multi-lingual retrievers on Hindi-BEIR provides important insights and a clear need for further research. We believe that Hindi-BEIR would greatly help in assessing retriever models on a standardized benchmark with focus on task and domain adaptability. We also envision that  Hindi-BEIR release would steer focused research towards developing robust retrievers models in Hindi and more indic languages in future.

\section{Limitations and Future Work}

The Hindi-BEIR Benchmark represents a foundational step towards creating a diverse benchmark for retrieval models in Hindi. Despite our efforts to ensure a high-quality benchmark, we acknowledge certain limitations in the current datasets.

Firstly, while we have covered eight domains, this scope might not fully encapsulate the breadth of tasks where retrieval models are crucial and need evaluation. In future work, we plan to expand the number of domains included in the Hindi-BEIR Benchmark. This expansion will encompass additional domains such as Law and Medicine, among others, to provide a more comprehensive evaluation framework.

Additionally, we aim to broaden the benchmark to include datasets from more low-resource languages, thereby enhancing its scope. We believe that a unified benchmark encompassing multiple low-resource languages will stimulate research aimed at developing superior retrieval models for these languages.

The performance of existing models on Hindi datasets, when compared to their performance on English datasets in similar domains, highlights a clear discrepancy. This underperformance underscores the necessity of developing retrieval models that can handle Hindi and English texts with equal efficacy. Addressing this challenge is something that we want to explore our future research endeavors.

By tackling these limitations and expanding the benchmark, we aim to drive advancements in the development of retrieval models that perform robustly across diverse languages and domains.
\bibliography{custom}

\begin{thebibliography}{36}
\providecommand{\natexlab}[1]{#1}

\bibitem[{Artetxe and Schwenk(2019)}]{Artetxe_2019}
Mikel Artetxe and Holger Schwenk. 2019.
\newblock \href {https://doi.org/10.1162/tacl_a_00288} {Massively multilingual sentence embeddings for zero-shot cross-lingual transfer and beyond}.
\newblock \emph{Transactions of the Association for Computational Linguistics}, 7:597–610.

\bibitem[{Bonifacio et~al.(2022)Bonifacio, Jeronymo, Abonizio, Campiotti, Fadaee, Lotufo, and Nogueira}]{bonifacio2022mmarco}
Luiz Bonifacio, Vitor Jeronymo, Hugo~Queiroz Abonizio, Israel Campiotti, Marzieh Fadaee, Roberto Lotufo, and Rodrigo Nogueira. 2022.
\newblock \href {https://arxiv.org/abs/2108.13897} {mmarco: A multilingual version of the ms marco passage ranking dataset}.
\newblock \emph{Preprint}, arXiv:2108.13897.

\bibitem[{Chen et~al.(2024)Chen, Xiao, Zhang, Luo, Lian, and Liu}]{chen2024bge}
Jianlv Chen, Shitao Xiao, Peitian Zhang, Kun Luo, Defu Lian, and Zheng Liu. 2024.
\newblock \href {https://arxiv.org/abs/2402.03216} {Bge m3-embedding: Multi-lingual, multi-functionality, multi-granularity text embeddings through self-knowledge distillation}.
\newblock \emph{Preprint}, arXiv:2402.03216.

\bibitem[{Cohan et~al.(2020)Cohan, Feldman, Beltagy, Downey, and Weld}]{cohan-etal-2020-specter}
Arman Cohan, Sergey Feldman, Iz~Beltagy, Doug Downey, and Daniel Weld. 2020.
\newblock \href {https://doi.org/10.18653/v1/2020.acl-main.207} {{SPECTER}: Document-level representation learning using citation-informed transformers}.
\newblock In \emph{Proceedings of the 58th Annual Meeting of the Association for Computational Linguistics}, pages 2270--2282, Online. Association for Computational Linguistics.

\bibitem[{Doddapaneni et~al.(2023)Doddapaneni, Aralikatte, Ramesh, Goyal, Khapra, Kunchukuttan, and Kumar}]{doddapaneni2023leaving}
Sumanth Doddapaneni, Rahul Aralikatte, Gowtham Ramesh, Shreya Goyal, Mitesh~M. Khapra, Anoop Kunchukuttan, and Pratyush Kumar. 2023.
\newblock \href {https://arxiv.org/abs/2212.05409} {Towards leaving no indic language behind: Building monolingual corpora, benchmark and models for indic languages}.
\newblock \emph{Preprint}, arXiv:2212.05409.

\bibitem[{Feng et~al.(2022)Feng, Yang, Cer, Arivazhagan, and Wang}]{feng2022languageagnostic}
Fangxiaoyu Feng, Yinfei Yang, Daniel Cer, Naveen Arivazhagan, and Wei Wang. 2022.
\newblock \href {https://arxiv.org/abs/2007.01852} {Language-agnostic bert sentence embedding}.
\newblock \emph{Preprint}, arXiv:2007.01852.

\bibitem[{Gala et~al.(2023)Gala, Chitale, Raghavan, Gumma, Doddapaneni, M, Nawale, Sujatha, Puduppully, Raghavan, Kumar, Khapra, Dabre, and Kunchukuttan}]{gala2023indictrans}
Jay Gala, Pranjal~A Chitale, A~K Raghavan, Varun Gumma, Sumanth Doddapaneni, Aswanth~Kumar M, Janki~Atul Nawale, Anupama Sujatha, Ratish Puduppully, Vivek Raghavan, Pratyush Kumar, Mitesh~M Khapra, Raj Dabre, and Anoop Kunchukuttan. 2023.
\newblock \href {https://openreview.net/forum?id=vfT4YuzAYA} {Indictrans2: Towards high-quality and accessible machine translation models for all 22 scheduled indian languages}.
\newblock \emph{Transactions on Machine Learning Research}.

\bibitem[{Ganguly et~al.(2021)Ganguly, Gangopadhyay, Mitra, and Majumder}]{DBLP:conf/fire/2021}
Debasis Ganguly, Surupendu Gangopadhyay, Mandar Mitra, and Prasenjit Majumder, editors. 2021.
\newblock \href {https://doi.org/10.1145/3503162} {\emph{{FIRE} 2021: Forum for Information Retrieval Evaluation, Virtual Event, India, December 13 - 17, 2021}}. {ACM}.

\bibitem[{Ganguly et~al.(2022)Ganguly, Gangopadhyay, Mitra, and Majumder}]{DBLP:conf/fire/2022}
Debasis Ganguly, Surupendu Gangopadhyay, Mandar Mitra, and Prasenjit Majumder, editors. 2022.
\newblock \href {https://doi.org/10.1145/3574318} {\emph{Proceedings of the 14th Annual Meeting of the Forum for Information Retrieval Evaluation, {FIRE} 2022, Kolkata, India, December 9-13, 2022}}. {ACM}.

\bibitem[{Ganguly et~al.(2023)Ganguly, Majumdar, Mitra, Gupta, Gangopadhyay, and Majumder}]{DBLP:conf/fire/2023}
Debasis Ganguly, Srijoni Majumdar, Bhaskar Mitra, Parth Gupta, Surupendu Gangopadhyay, and Prasenjit Majumder, editors. 2023.
\newblock \href {https://doi.org/10.1145/3632754} {\emph{Proceedings of the 15th Annual Meeting of the Forum for Information Retrieval Evaluation, {FIRE} 2023, Panjim, India, December 15-18, 2023}}. {ACM}.

\bibitem[{Haq et~al.(2023)Haq, Sharma, and Bhattacharyya}]{haq2023indicirsuite}
Saiful Haq, Ashutosh Sharma, and Pushpak Bhattacharyya. 2023.
\newblock \href {https://arxiv.org/abs/2312.09508} {Indicirsuite: Multilingual dataset and neural information models for indian languages}.
\newblock \emph{Preprint}, arXiv:2312.09508.

\bibitem[{J{\"a}rvelin and Kek{\"a}l{\"a}inen(2002)}]{Jrvelin2002CumulatedGE}
Kalervo J{\"a}rvelin and Jaana Kek{\"a}l{\"a}inen. 2002.
\newblock \href {https://api.semanticscholar.org/CorpusID:1981391} {Cumulated gain-based evaluation of ir techniques}.
\newblock \emph{ACM Trans. Inf. Syst.}, 20:422--446.

\bibitem[{Lewis et~al.(2021)Lewis, Perez, Piktus, Petroni, Karpukhin, Goyal, Küttler, Lewis, tau Yih, Rocktäschel, Riedel, and Kiela}]{lewis2021retrievalaugmented}
Patrick Lewis, Ethan Perez, Aleksandra Piktus, Fabio Petroni, Vladimir Karpukhin, Naman Goyal, Heinrich Küttler, Mike Lewis, Wen tau Yih, Tim Rocktäschel, Sebastian Riedel, and Douwe Kiela. 2021.
\newblock \href {https://arxiv.org/abs/2005.11401} {Retrieval-augmented generation for knowledge-intensive nlp tasks}.
\newblock \emph{Preprint}, arXiv:2005.11401.

\bibitem[{Lin(2004)}]{lin-2004-rouge}
Chin-Yew Lin. 2004.
\newblock \href {https://aclanthology.org/W04-1013} {{ROUGE}: A package for automatic evaluation of summaries}.
\newblock In \emph{Text Summarization Branches Out}, pages 74--81, Barcelona, Spain. Association for Computational Linguistics.

\bibitem[{Maia et~al.(2018)Maia, Handschuh, Freitas, Davis, McDermott, Zarrouk, and Balahur}]{10.1145/3184558.3192301}
Macedo Maia, Siegfried Handschuh, Andr\'{e} Freitas, Brian Davis, Ross McDermott, Manel Zarrouk, and Alexandra Balahur. 2018.
\newblock \href {https://doi.org/10.1145/3184558.3192301} {Www'18 open challenge: Financial opinion mining and question answering}.
\newblock In \emph{Companion Proceedings of the The Web Conference 2018}, WWW '18, page 1941–1942, Republic and Canton of Geneva, CHE. International World Wide Web Conferences Steering Committee.

\bibitem[{Majumder et~al.(2013{\natexlab{a}})Majumder, Mitra, Agrawal, and Mehta}]{DBLP:conf/fire/2013}
Prasenjit Majumder, Mandar Mitra, Madhulika Agrawal, and Parth Mehta, editors. 2013{\natexlab{a}}.
\newblock \href {http://dl.acm.org/citation.cfm?id=2701336} {\emph{Proceedings of the 5th 2013 Forum on Information Retrieval Evaluation, {FIRE} '13, New Delhi, India, December 4-6, 2013}}. {ACM}.

\bibitem[{Majumder et~al.(2014)Majumder, Mitra, Agrawal, and Mehta}]{DBLP:conf/fire/2014}
Prasenjit Majumder, Mandar Mitra, Madhulika Agrawal, and Parth Mehta, editors. 2014.
\newblock \href {http://dl.acm.org/citation.cfm?id=2824864} {\emph{Proceedings of the Forum for Information Retrieval Evaluation, {FIRE} 2014, Bangalore, India, December 5-7, 2014}}. {ACM}.

\bibitem[{Majumder et~al.(2015)Majumder, Mitra, Agrawal, and Mehta}]{DBLP:conf/fire/2015}
Prasenjit Majumder, Mandar Mitra, Madhulika Agrawal, and Parth Mehta, editors. 2015.
\newblock \href {https://doi.org/10.1145/2838706} {\emph{Proceedings of the 7th Forum for Information Retrieval Evaluation, {FIRE} 2015, Gandhinagar, India, December 4-6, 2015}}. {ACM}.

\bibitem[{Majumder et~al.(2013{\natexlab{b}})Majumder, Mitra, Bhattacharyya, Subramaniam, Contractor, and Rosso}]{DBLP:conf/fire/2011}
Prasenjit Majumder, Mandar Mitra, Pushpak Bhattacharyya, L.~Venkata Subramaniam, Danish Contractor, and Paolo Rosso, editors. 2013{\natexlab{b}}.
\newblock \href {https://doi.org/10.1007/978-3-642-40087-2} {\emph{Multilingual Information Access in South Asian Languages - Second International Workshop, {FIRE} 2010, Gandhinagar, India, February 19-21, 2010 and Third International Workshop, {FIRE} 2011, Bombay, India, December 2-4, 2011, Revised Selected Papers}}, volume 7536 of \emph{Lecture Notes in Computer Science}. Springer.

\bibitem[{Majumder et~al.(2019)Majumder, Mitra, Gangopadhyay, and Mehta}]{DBLP:conf/fire/2019}
Prasenjit Majumder, Mandar Mitra, Surupendu Gangopadhyay, and Parth Mehta, editors. 2019.
\newblock \href {https://doi.org/10.1145/3368567} {\emph{{FIRE} '19: Forum for Information Retrieval Evaluation, Kolkata, India, December, 2019}}. {ACM}.

\bibitem[{Majumder et~al.(2020)Majumder, Mitra, Gangopadhyay, and Mehta}]{DBLP:conf/fire/2020}
Prasenjit Majumder, Mandar Mitra, Surupendu Gangopadhyay, and Parth Mehta, editors. 2020.
\newblock \href {https://doi.org/10.1145/3441501} {\emph{{FIRE} 2020: Forum for Information Retrieval Evaluation, Hyderabad, India, December 16-20, 2020}}. {ACM}.

\bibitem[{Majumder et~al.(2016)Majumder, Mitra, Sankhavara, and Mehta}]{DBLP:conf/fire/2016}
Prasenjit Majumder, Mandar Mitra, Jainisha Sankhavara, and Parth Mehta, editors. 2016.
\newblock \href {https://doi.org/10.1145/3015157} {\emph{Proceedings of the 8th annual meeting of the Forum on Information Retrieval Evaluation, {FIRE} 2016, Kolkata, India, December 8-10, 2016}}. {ACM}.

\bibitem[{Majumder et~al.(2017)Majumder, Mitra, Sankhavara, and Mehta}]{DBLP:conf/fire/2017}
Prasenjit Majumder, Mandar Mitra, Jainisha Sankhavara, and Parth Mehta, editors. 2017.
\newblock \href {http://dl.acm.org/citation.cfm?id=3158354} {\emph{Proceedings of the 9th annual meeting of the Forum for Information Retrieval Evaluation, {FIRE} 2017, Bangalore, India, December 8-10, 2017}}. {ACM}.

\bibitem[{Majumder et~al.(2018)Majumder, Mitra, Sankhavara, and Mehta}]{DBLP:conf/fire/2018}
Prasenjit Majumder, Mandar Mitra, Jainisha Sankhavara, and Parth Mehta, editors. 2018.
\newblock \href {https://doi.org/10.1145/3293339} {\emph{Proceedings of the 10th annual meeting of the Forum for Information Retrieval Evaluation, {FIRE} 2018, Gandhinagar, India, December 06-09, 2018}}. {ACM}.

\bibitem[{Nguyen et~al.(2016)Nguyen, Rosenberg, Song, Gao, Tiwary, Majumder, and Deng}]{DBLP:journals/corr/NguyenRSGTMD16}
Tri Nguyen, Mir Rosenberg, Xia Song, Jianfeng Gao, Saurabh Tiwary, Rangan Majumder, and Li~Deng. 2016.
\newblock \href {https://arxiv.org/abs/1611.09268} {{MS} {MARCO:} {A} human generated machine reading comprehension dataset}.
\newblock \emph{CoRR}, abs/1611.09268.

\bibitem[{Popovi{\'c}(2017)}]{popovic-2017-chrf}
Maja Popovi{\'c}. 2017.
\newblock \href {https://doi.org/10.18653/v1/W17-4770} {chr{F}++: words helping character n-grams}.
\newblock In \emph{Proceedings of the Second Conference on Machine Translation}, pages 612--618, Copenhagen, Denmark. Association for Computational Linguistics.

\bibitem[{Thakur et~al.(2021)Thakur, Reimers, Rücklé, Srivastava, and Gurevych}]{thakur2021beir}
Nandan Thakur, Nils Reimers, Andreas Rücklé, Abhishek Srivastava, and Iryna Gurevych. 2021.
\newblock \href {https://arxiv.org/abs/2104.08663} {Beir: A heterogenous benchmark for zero-shot evaluation of information retrieval models}.
\newblock \emph{Preprint}, arXiv:2104.08663.

\bibitem[{van~den Oord et~al.(2019)van~den Oord, Li, and Vinyals}]{oord2019representation}
Aaron van~den Oord, Yazhe Li, and Oriol Vinyals. 2019.
\newblock \href {https://arxiv.org/abs/1807.03748} {Representation learning with contrastive predictive coding}.
\newblock \emph{Preprint}, arXiv:1807.03748.

\bibitem[{Voorhees et~al.(2021)Voorhees, Alam, Bedrick, Demner-Fushman, Hersh, Lo, Roberts, Soboroff, and Wang}]{10.1145/3451964.3451965}
Ellen Voorhees, Tasmeer Alam, Steven Bedrick, Dina Demner-Fushman, William~R. Hersh, Kyle Lo, Kirk Roberts, Ian Soboroff, and Lucy~Lu Wang. 2021.
\newblock \href {https://doi.org/10.1145/3451964.3451965} {Trec-covid: constructing a pandemic information retrieval test collection}.
\newblock \emph{SIGIR Forum}, 54(1).

\bibitem[{Wachsmuth et~al.(2017)Wachsmuth, Potthast, Al-Khatib, Ajjour, Puschmann, Qu, Dorsch, Morari, Bevendorff, and Stein}]{wachsmuth-etal-2017-building}
Henning Wachsmuth, Martin Potthast, Khalid Al-Khatib, Yamen Ajjour, Jana Puschmann, Jiani Qu, Jonas Dorsch, Viorel Morari, Janek Bevendorff, and Benno Stein. 2017.
\newblock \href {https://doi.org/10.18653/v1/W17-5106} {Building an argument search engine for the web}.
\newblock In \emph{Proceedings of the 4th Workshop on Argument Mining}, pages 49--59, Copenhagen, Denmark. Association for Computational Linguistics.

\bibitem[{Wachsmuth et~al.(2018)Wachsmuth, Syed, and Stein}]{wachsmuth-etal-2018-retrieval}
Henning Wachsmuth, Shahbaz Syed, and Benno Stein. 2018.
\newblock \href {https://doi.org/10.18653/v1/P18-1023} {Retrieval of the best counterargument without prior topic knowledge}.
\newblock In \emph{Proceedings of the 56th Annual Meeting of the Association for Computational Linguistics (Volume 1: Long Papers)}, pages 241--251, Melbourne, Australia. Association for Computational Linguistics.

\bibitem[{Wang et~al.(2024{\natexlab{a}})Wang, Yang, Huang, Jiao, Yang, Jiang, Majumder, and Wei}]{wang2024text}
Liang Wang, Nan Yang, Xiaolong Huang, Binxing Jiao, Linjun Yang, Daxin Jiang, Rangan Majumder, and Furu Wei. 2024{\natexlab{a}}.
\newblock \href {https://arxiv.org/abs/2212.03533} {Text embeddings by weakly-supervised contrastive pre-training}.
\newblock \emph{Preprint}, arXiv:2212.03533.

\bibitem[{Wang et~al.(2024{\natexlab{b}})Wang, Yang, Huang, Yang, Majumder, and Wei}]{wang2024multilingual}
Liang Wang, Nan Yang, Xiaolong Huang, Linjun Yang, Rangan Majumder, and Furu Wei. 2024{\natexlab{b}}.
\newblock \href {https://arxiv.org/abs/2402.05672} {Multilingual e5 text embeddings: A technical report}.
\newblock \emph{Preprint}, arXiv:2402.05672.

\bibitem[{Zhang et~al.(2021)Zhang, Ma, Shi, and Lin}]{zhang-etal-2021-mr}
Xinyu Zhang, Xueguang Ma, Peng Shi, and Jimmy Lin. 2021.
\newblock \href {https://doi.org/10.18653/v1/2021.mrl-1.12} {Mr. {T}y{D}i: A multi-lingual benchmark for dense retrieval}.
\newblock In \emph{Proceedings of the 1st Workshop on Multilingual Representation Learning}, pages 127--137, Punta Cana, Dominican Republic. Association for Computational Linguistics.

\bibitem[{Zhang et~al.(2022)Zhang, Thakur, Ogundepo, Kamalloo, Alfonso-Hermelo, Li, Liu, Rezagholizadeh, and Lin}]{zhang2022making}
Xinyu Zhang, Nandan Thakur, Odunayo Ogundepo, Ehsan Kamalloo, David Alfonso-Hermelo, Xiaoguang Li, Qun Liu, Mehdi Rezagholizadeh, and Jimmy Lin. 2022.
\newblock \href {https://arxiv.org/abs/2210.09984} {Making a miracl: Multilingual information retrieval across a continuum of languages}.
\newblock \emph{Preprint}, arXiv:2210.09984.

\bibitem[{Zhang et~al.(2023)Zhang, Thakur, Ogundepo, Kamalloo, Alfonso-Hermelo, Li, Liu, Rezagholizadeh, and Lin}]{10.1162/tacl_a_00595}
Xinyu Zhang, Nandan Thakur, Odunayo Ogundepo, Ehsan Kamalloo, David Alfonso-Hermelo, Xiaoguang Li, Qun Liu, Mehdi Rezagholizadeh, and Jimmy Lin. 2023.
\newblock \href {https://doi.org/10.1162/tacl_a_00595} {{MIRACL: A Multilingual Retrieval Dataset Covering 18 Diverse Languages}}.
\newblock \emph{Transactions of the Association for Computational Linguistics}, 11:1114--1131.

\end{thebibliography}

\appendix

\section{Appendix}
\label{sec:appendix}

\subsection{Dataset Description}
\label{subsec:dataset_desc}
\subsubsection{ArguAna}
\begin{enumerate}
\item \textbf{Task definition:} Derived from the work by \citeauthor{wachsmuth-etal-2018-retrieval} , the task is to retrieve the best counter argument, given an argument. Translations of arguments and counter-arguments from online debates constitute the corpus while the translations of arguments in the original test split, after going through the filtration process based on Chrf++ scores (refer to \ref{subsec:translation}), constitute the queries.

    \item \textbf{Domain :} Misc.
\end{enumerate}

\begin{figure}[hbt!]
    \centering
    \includegraphics[height=8cm,width=8cm]{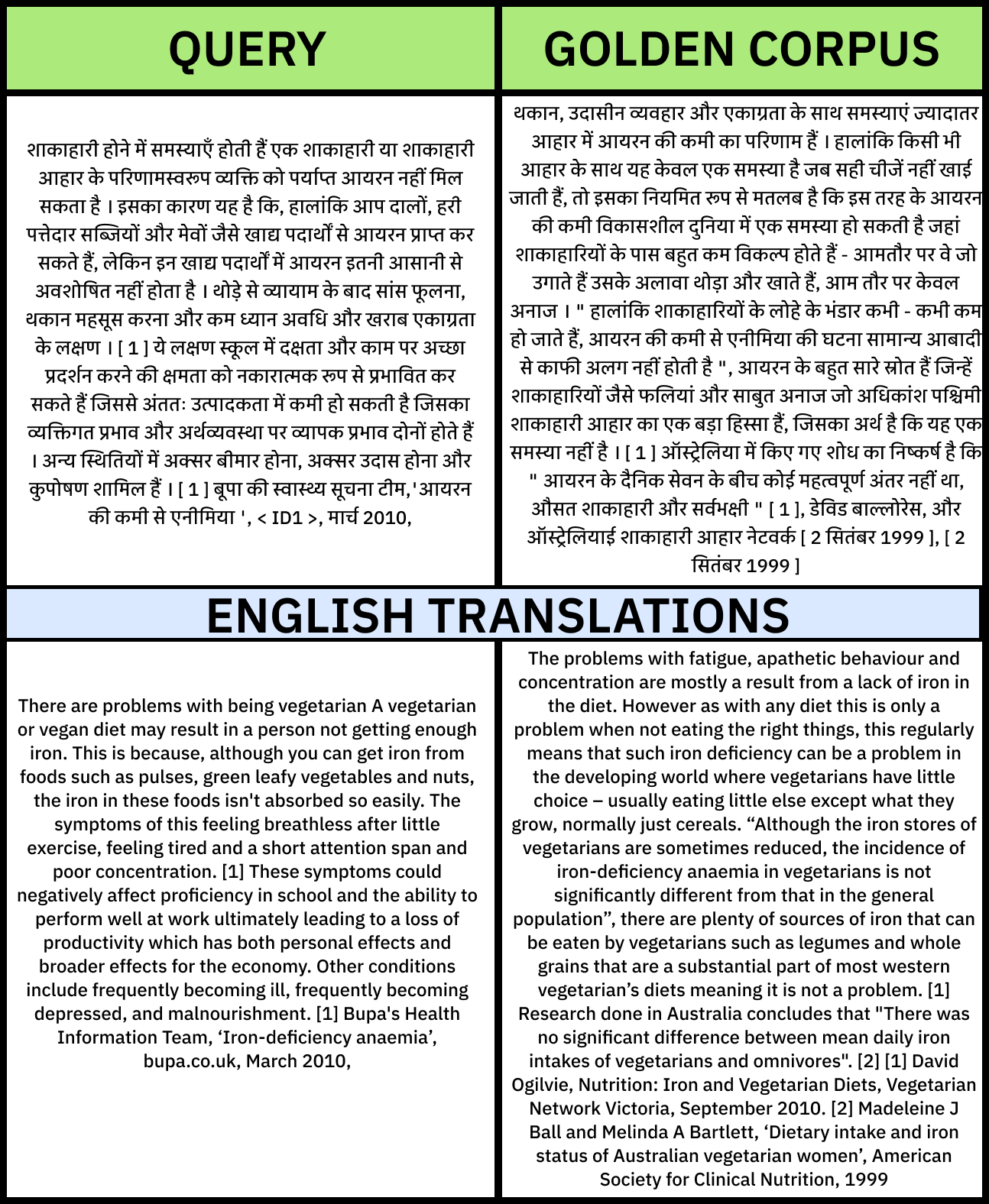}
    \caption{An example of a query with its corresponding golden corpus from the ArguAna Dataset}
    \label{fig:arguana_example}
\end{figure}

An example of query with its corresponding golden corpus has been provided in Figure \ref{fig:arguana_example}

\begin{figure}[hbt!]
    \centering
    \includegraphics[height=4cm,width=8cm]{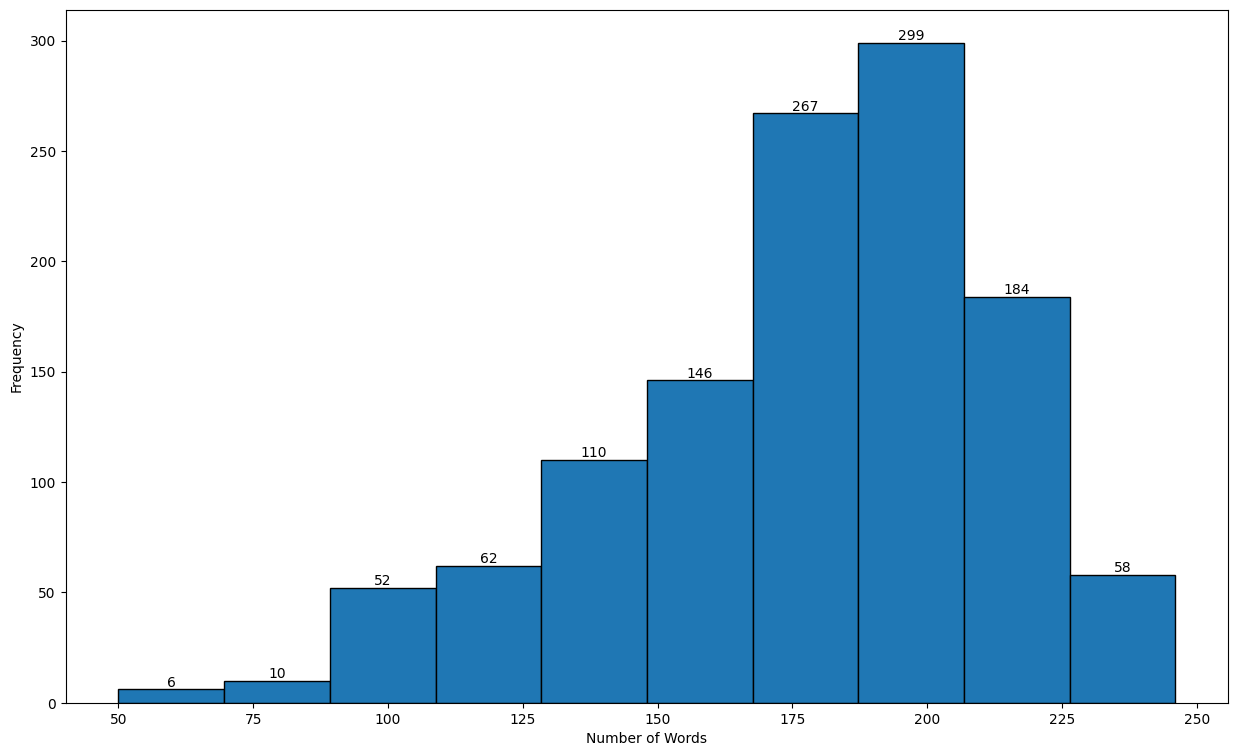}
    \caption{Distribution of the number of words in the queries of ArguAna Dataset}
    \label{fig:arguana_queries}
\end{figure}

\begin{figure}[hbt!]
    \centering
    \includegraphics[height=4cm,width=8cm]{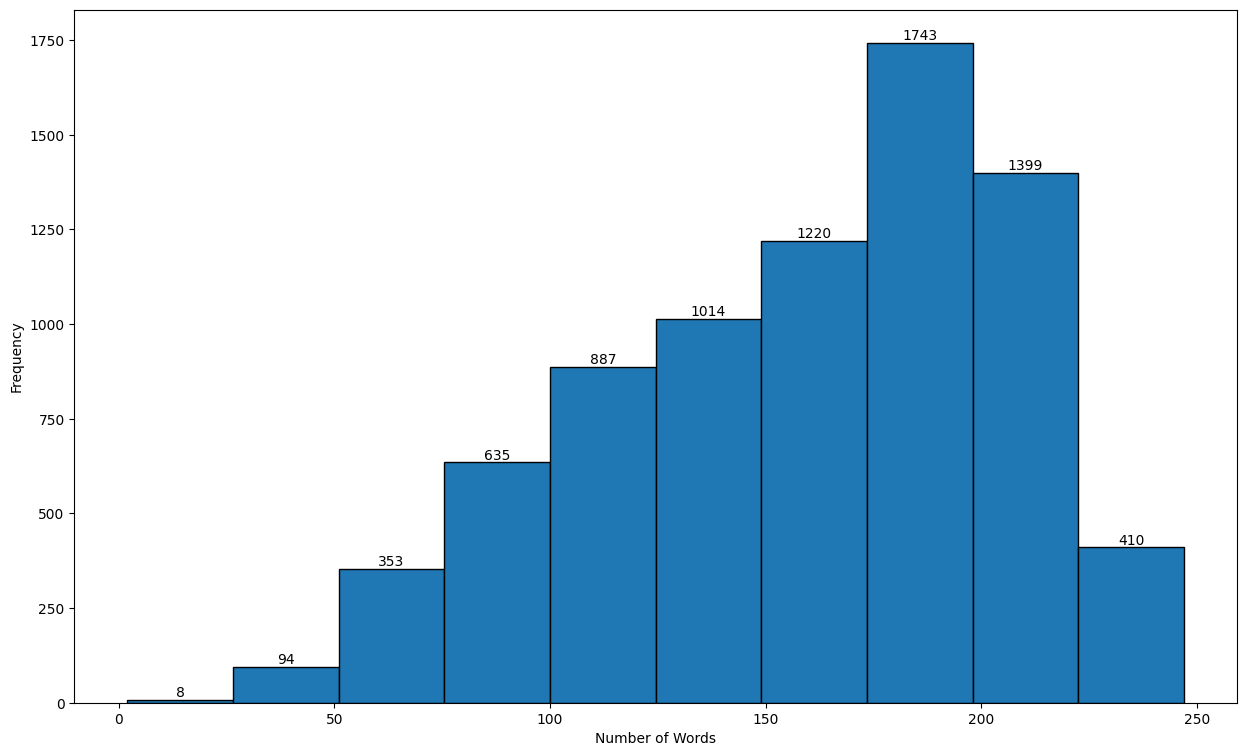}
    \caption{Distribution of Number of Words in corpus of ArguAna Dataset}
    \label{fig:arguana_corpus}
\end{figure}

Distribution of the number of words in the corpus and queries in ArguAna dataset has been shown in Figure \ref{fig:arguana_corpus} and Figure \ref{fig:arguana_queries} respectively.

\subsubsection{FiQA-2018}
\begin{enumerate}
\item \textbf{Task Definition: }It deals with Opinion-Based Question answering. Based on the works of \citeauthor{10.1145/3184558.3192301},translation of the financial data
extracted by crawling StackExchange posts under the Investment topic from 2009-2017, after passing through filteration processes mentioned in \ref{subsec:translation}, acts as the corpus. While translations from the original training split acts as the queries.

    \item \textbf{Domain :} Finance
\end{enumerate}

\begin{figure}[hbt!]
    \centering
    \includegraphics[height=6cm,width=8cm]{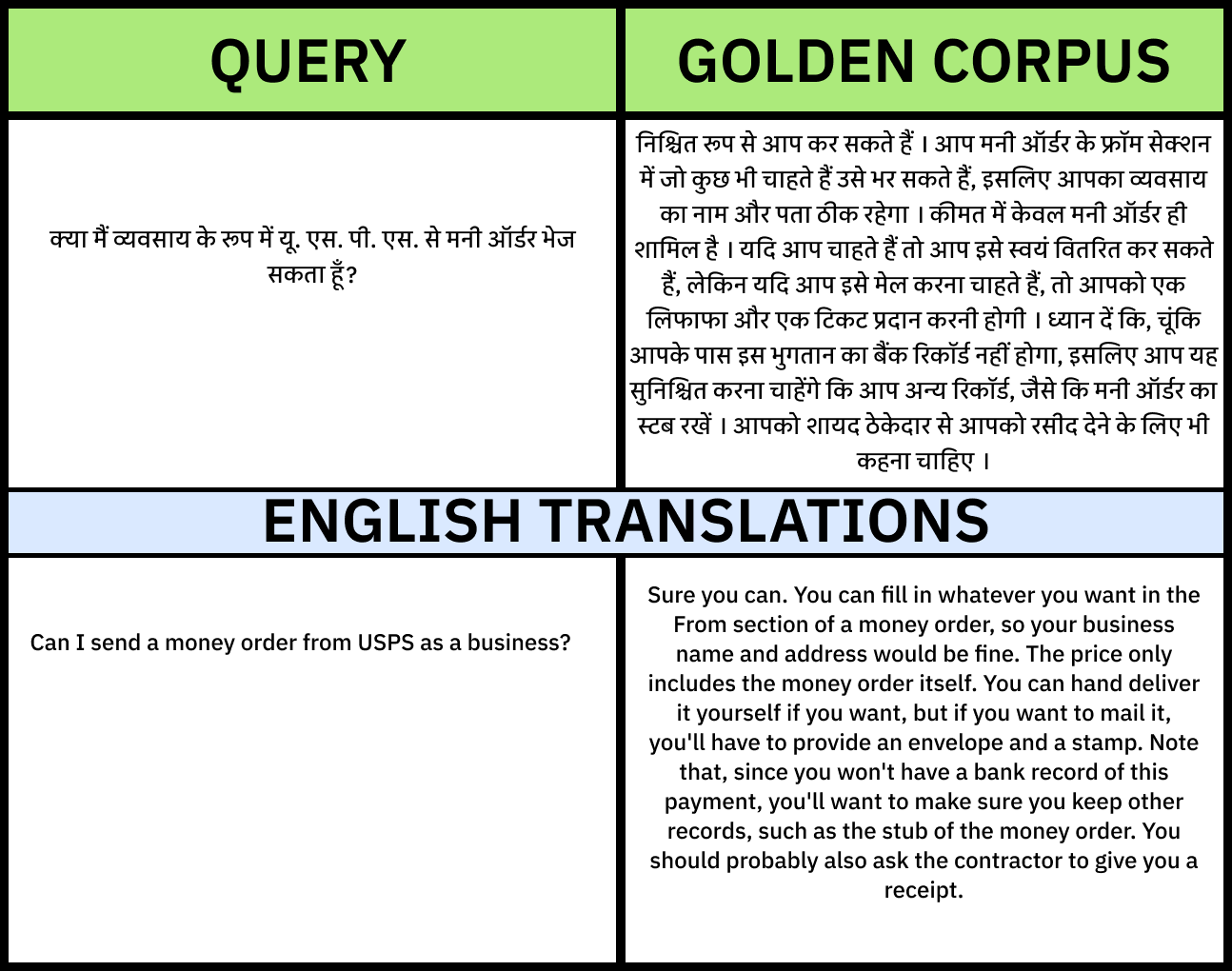}
    \caption{An example of a query with its corresponding golden corpus from the FiQA-2018 Dataset}
    \label{fig:fiqa_example}
\end{figure}

An example of query with its corresponding golden corpus from the FiQA-2018 dataset has been provided in Figure \ref{fig:fiqa_example}

\begin{figure}[hbt!]
    \centering
    \includegraphics[height=4cm,width=8cm]{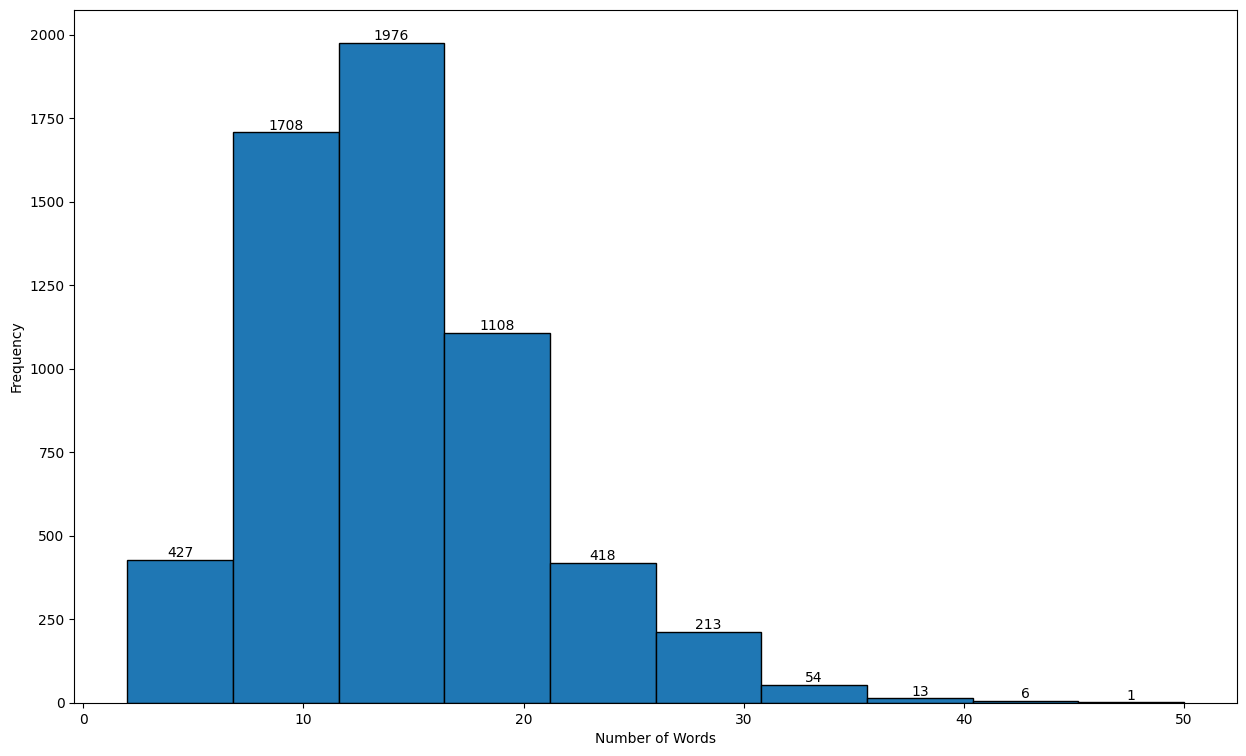}
    \caption{Distribution of the number of words in the queries of FiQA-2018 Dataset}
    \label{fig:fiqa_queries}
\end{figure}

\begin{figure}[hbt!]
    \centering
    \includegraphics[height=4cm,width=8cm]{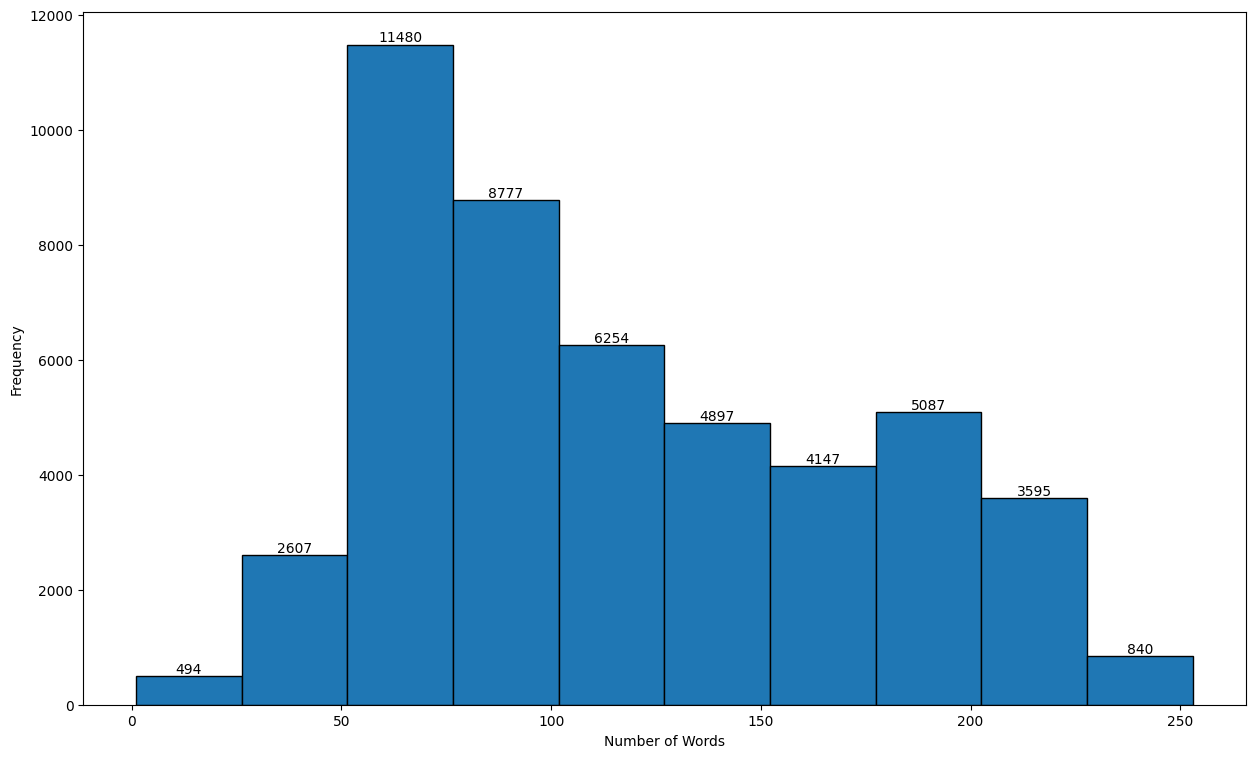}
    \caption{Distribution of Number of Words in corpus of FiQA-2018 Dataset}
    \label{fig:fiqa_corpus}
\end{figure}

Distribution of the number of words in the corpus and queries in FiQA-2018 dataset has been shown in Figure \ref{fig:fiqa_corpus} and Figure \ref{fig:fiqa_queries} respectively.

\subsubsection{TREC-COVID}
\begin{enumerate}
\item \textbf{Task Definition: }\citeauthor{10.1145/3451964.3451965} introduced the original TREC-COVID dataset which is an ad-hoc seach challenge based on CORD-19 dataset containing articles about the COVID-19 Pandemic. The translated and filtered version of the CORD-19 Dataset constitutes the corpus while the final cumulative judgements with query descriptions from the original task are the queries in the TREC-COVID Dataset.

    \item \textbf{Domain :} Bio-Medical
\end{enumerate}

\begin{figure}[hbt!]
    \centering
    \includegraphics[height=8.5cm,width=8cm]{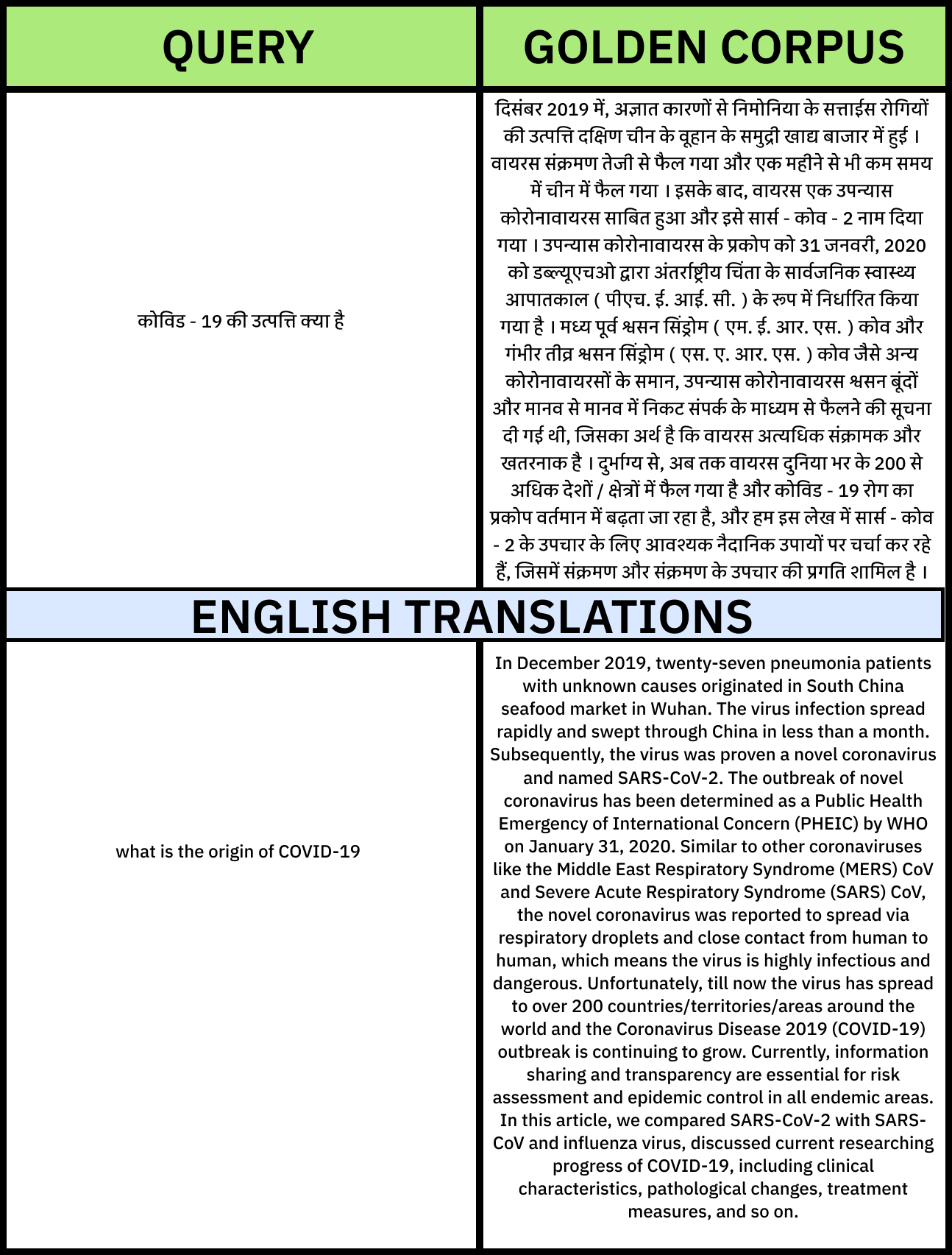}
    \caption{An example of a query with its corresponding golden corpus from the TREC-COVID Dataset}
    \label{fig:trec_example}
\end{figure}

An example of query with its corresponding golden corpus from the TREC-COVID dataset has been provided in Figure \ref{fig:trec_example}

\begin{figure}[hbt!]
    \centering
    \includegraphics[height=4cm,width=8cm]{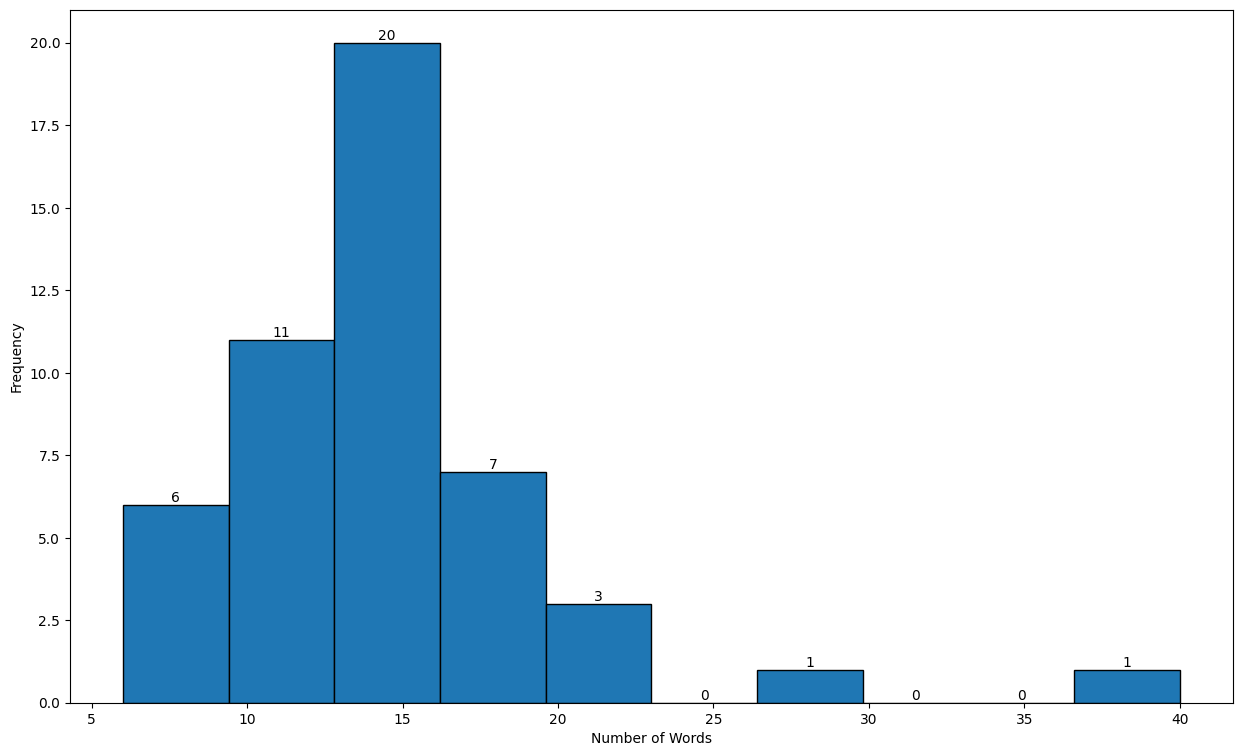}
    \caption{Distribution of the number of words in the queries of TREC-COVID Dataset}
    \label{fig:trec_queries}
\end{figure}

\begin{figure}[hbt!]
    \centering
    \includegraphics[height=4cm,width=8cm]{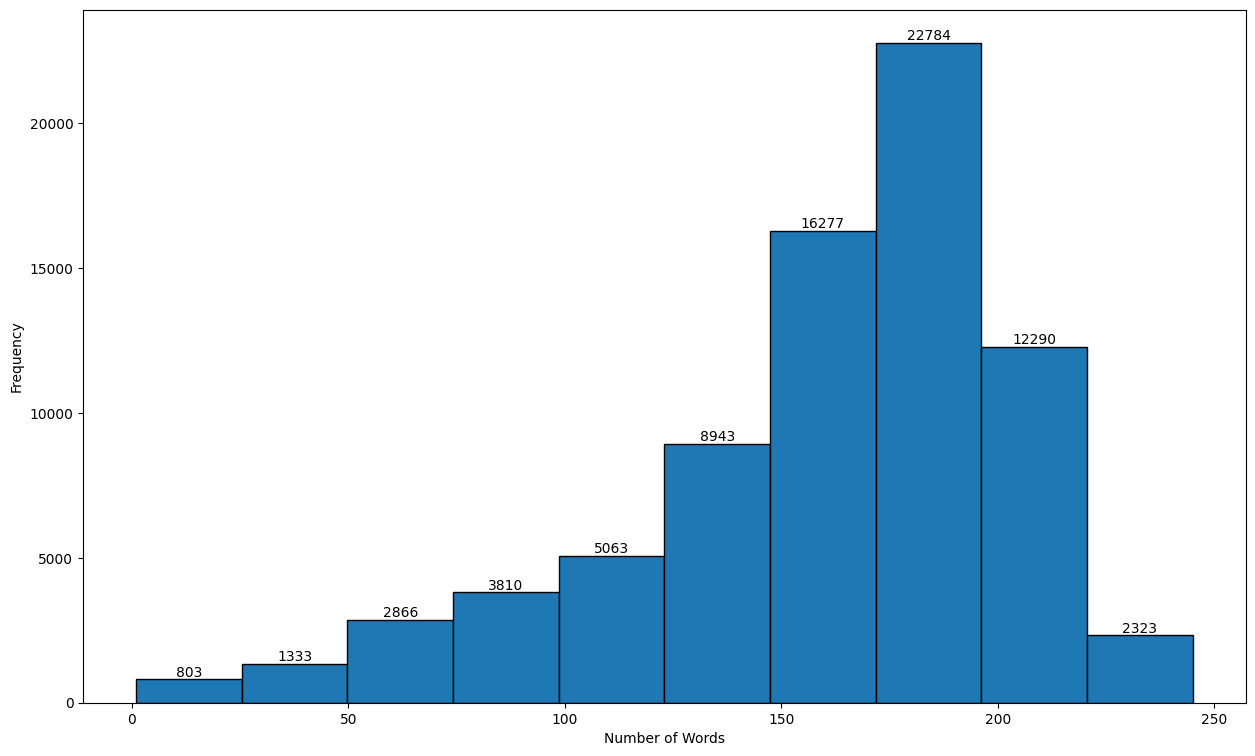}
    \caption{Distribution of Number of Words in corpus of TREC-COVID Dataset}
    \label{fig:trec_corpus}
\end{figure}

Distribution of the number of words in the corpus and queries in TREC-COVID dataset has been shown in Figure \ref{fig:trec_corpus} and Figure \ref{fig:trec_queries} respectively.

\subsubsection{SCIDOCS}
\begin{enumerate}
\item \textbf{Task Definition: } Inspired by \citeauthor{cohan-etal-2020-specter}, in this task we expect the model to retrieve cited papers for a given scientific paper abstract as input. The corpus contains about 22k translated and filtered scientific paper abstracts and 850 translated paper titles as queries.

    \item \textbf{Domain :} Scientific
\end{enumerate} 

\begin{figure}[hbt!]
    \centering
    \includegraphics[height=6cm,width=8cm]{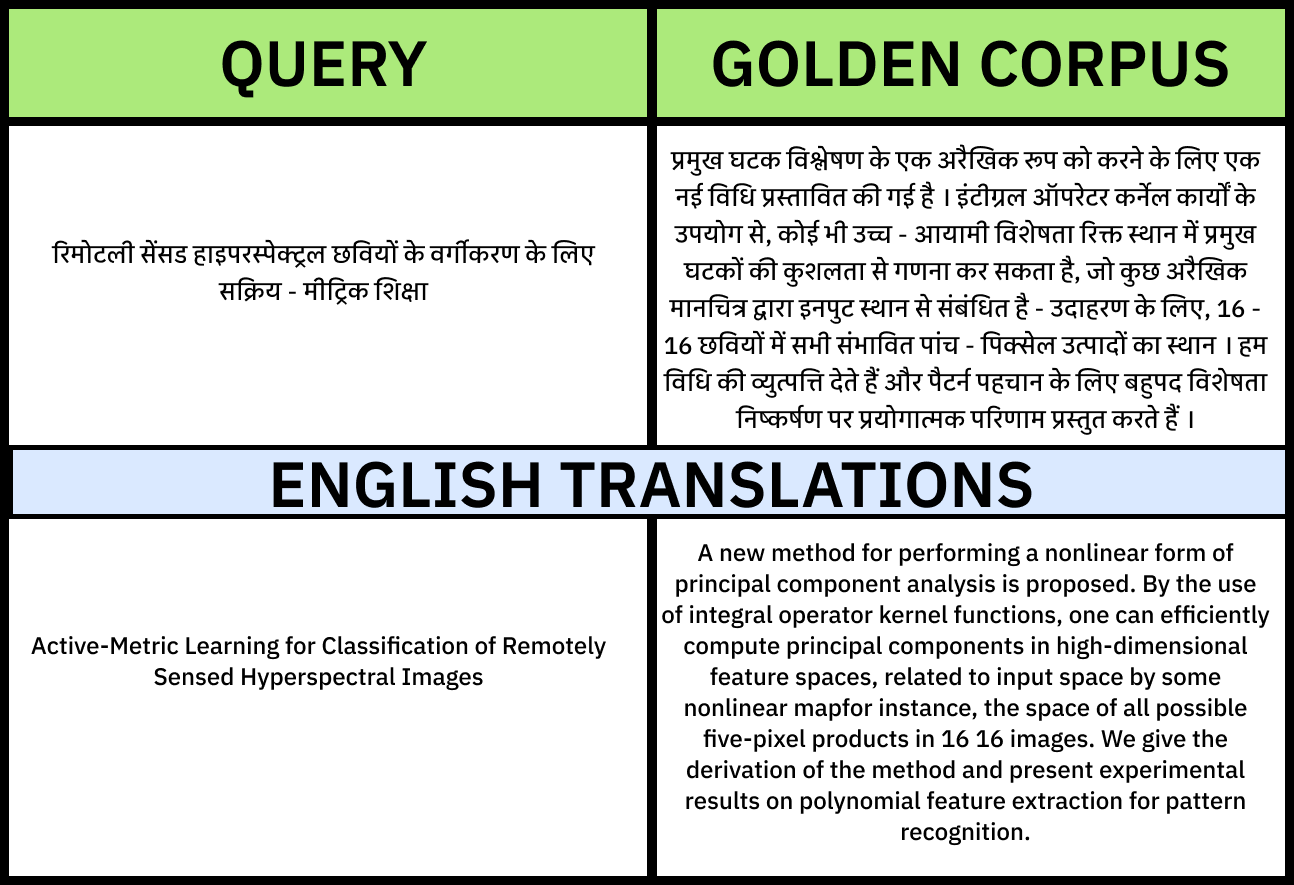}
    \caption{An example of a query with its corresponding golden corpus from the SCIDOCS Dataset}
    \label{fig:scidocs_example}
\end{figure}

An example of query with its corresponding golden corpus from the SCIDOCS dataset has been provided in Figure \ref{fig:scidocs_example}

\begin{figure}[hbt!]
    \centering
    \includegraphics[height=4cm,width=8cm]{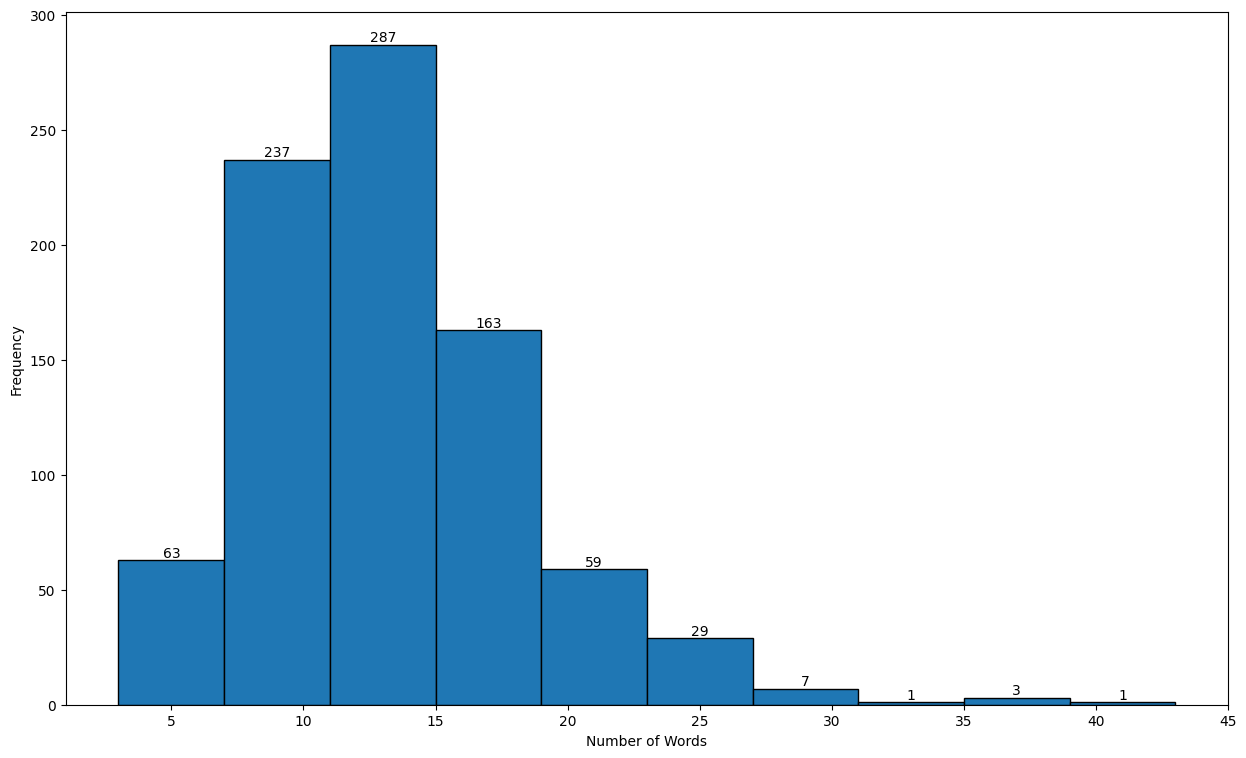}
    \caption{Distribution of the number of words in the queries of SCIDOCS Dataset}
    \label{fig:scidocs_queries}
\end{figure}

\begin{figure}[hbt!]
    \centering
    \includegraphics[height=4cm,width=8cm]{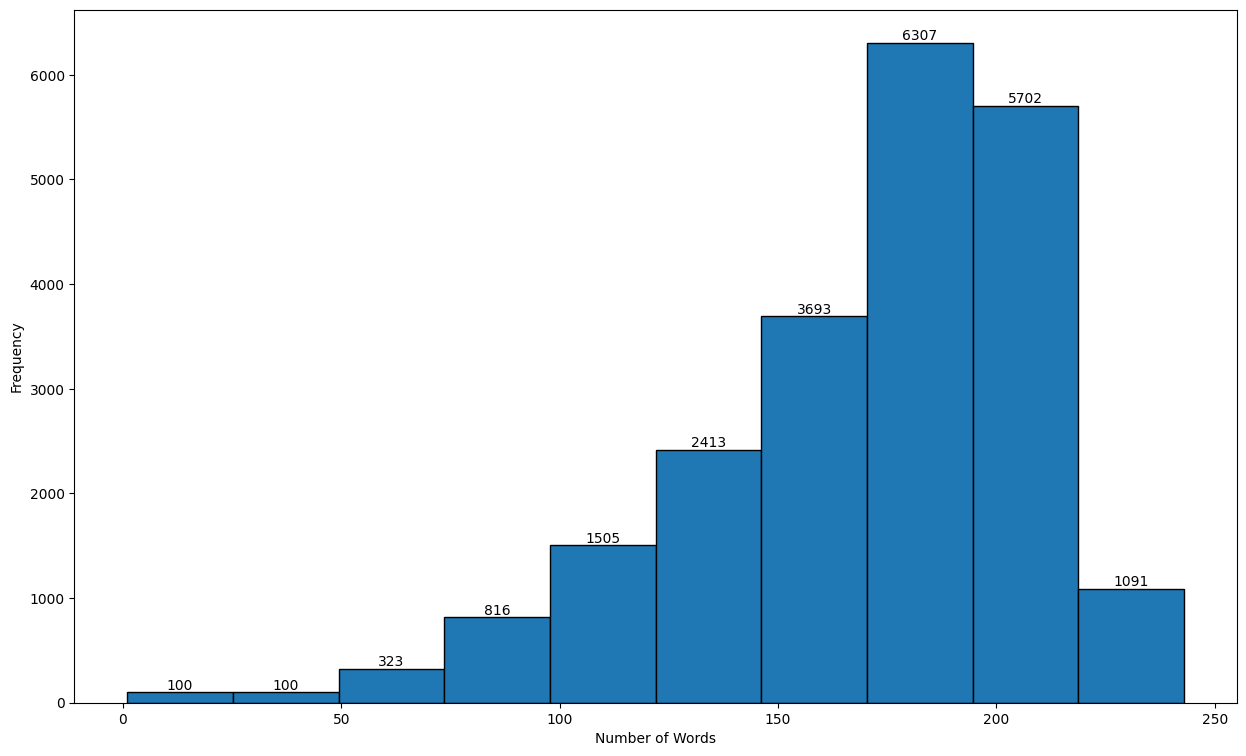}
    \caption{Distribution of Number of Words in corpus of SCIDOCS Dataset}
    \label{fig:scidocs_corpus}
\end{figure}

Distribution of the number of words in the corpus and queries in SCIDOCS dataset has been shown in Figure \ref{fig:scidocs_corpus} and Figure \ref{fig:scidocs_queries} respectively.
\subsubsection{SciFact}
\begin{enumerate}
\item \textbf{Task Definition: } This task involves verification of scientific claims given the abstract of scientific articles from recent literature. For this task the model is expected to retrieve relevant abstract which a given claim as input.

    \item \textbf{Domain :} Scientific
\end{enumerate} 

\begin{figure}[hbt!]
    \centering
    \includegraphics[height=8cm,width=8cm]{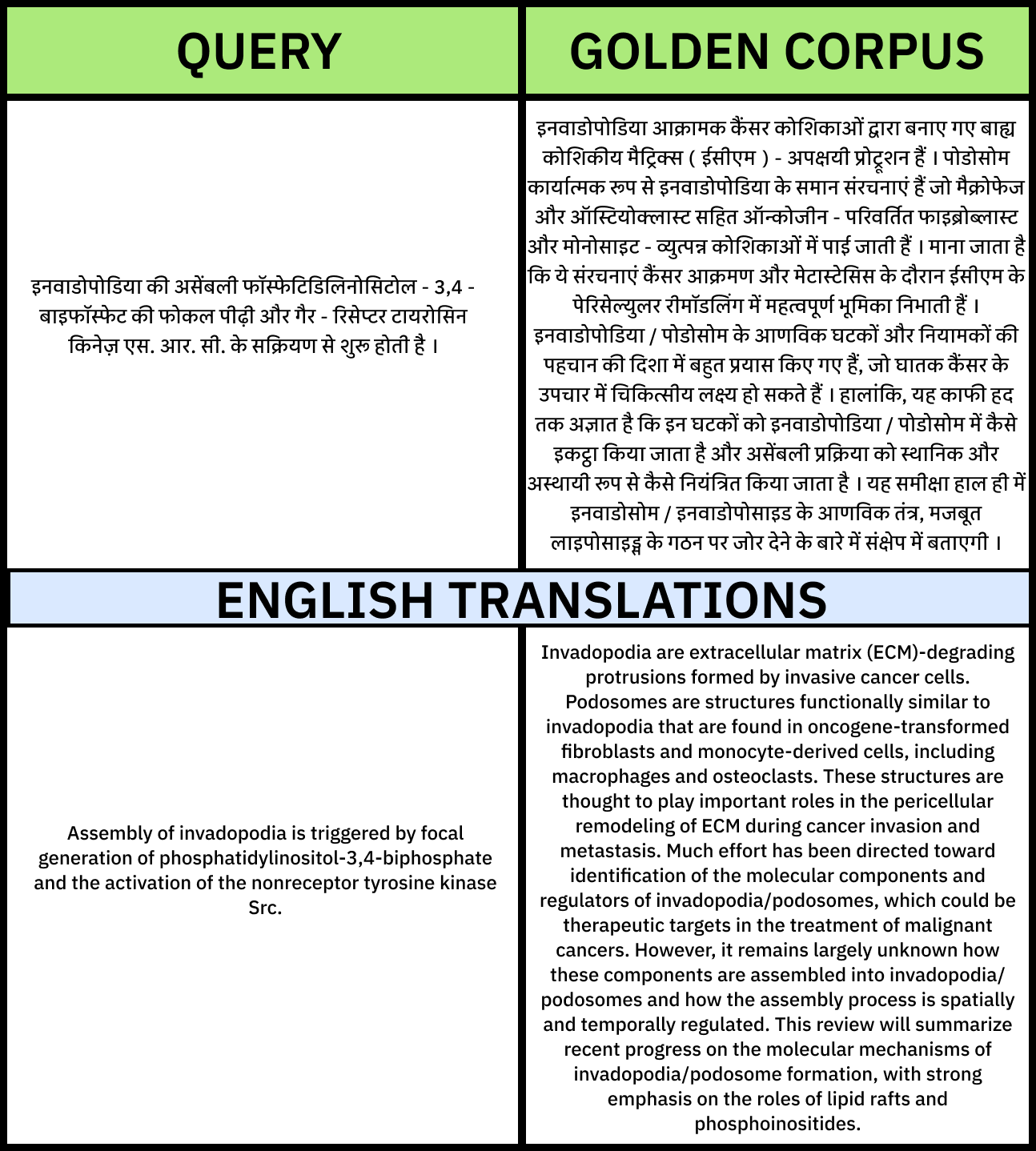}
    \caption{An example of a query with its corresponding golden corpus from the SciFact Dataset}
    \label{fig:scifact_example}
\end{figure}

An example of query with its corresponding golden corpus from the SciFact dataset has been provided in Figure \ref{fig:scifact_example}

\begin{figure}[hbt!]
    \centering
    \includegraphics[height=4cm,width=8cm]{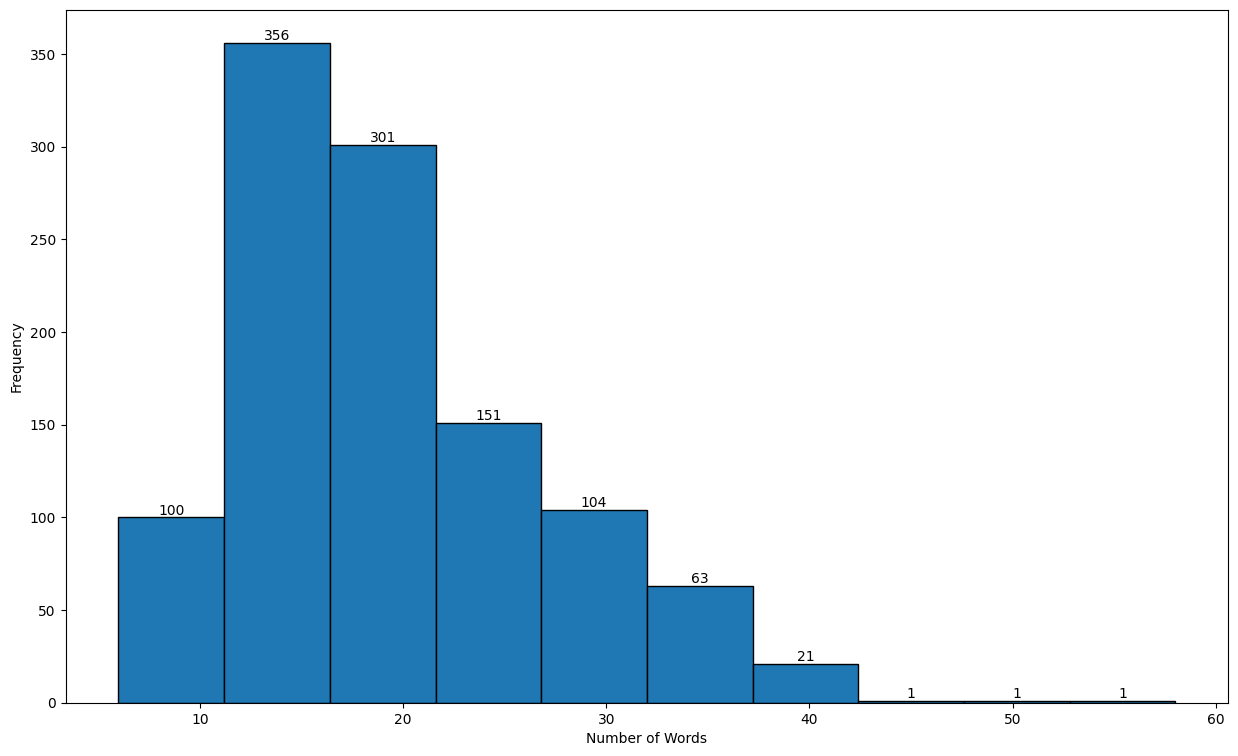}
    \caption{Distribution of the number of words in the queries of SciFact Dataset}
    \label{fig:scifact_queries}
\end{figure}

\begin{figure}[hbt!]
    \centering
    \includegraphics[height=4cm,width=8cm]{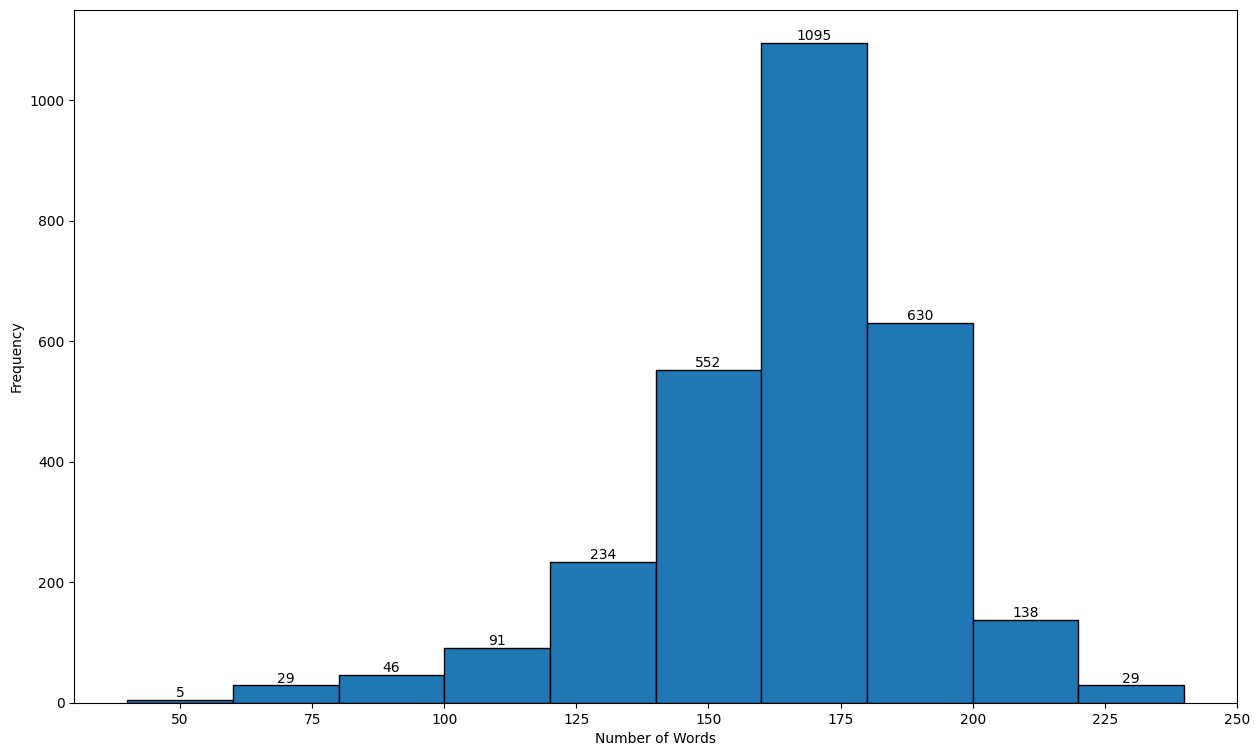}
    \caption{Distribution of Number of Words in corpus of SciFact Dataset}
    \label{fig:scifact_corpus}
\end{figure}

Distribution of the number of words in the corpus and queries in SciFact dataset has been shown in Figure \ref{fig:scifact_corpus} and Figure \ref{fig:scifact_queries} respectively.

\subsubsection{Touch\'e-2020}
\begin{enumerate}
\item \textbf{Task Definition: } Similar to ArguAna this task deals with this task deals with the retrieval of conversational arguments.  The  translated and filtered conclusion forms the title and premise for arguments present in args.me \cite{wachsmuth-etal-2017-building} constitutes the corpus. The translations of the Touch\'e-2020 task data are the queries.

    \item \textbf{Domain :} Miscellaneous
\end{enumerate} 

\begin{figure}[hbt!]
    \centering
    \includegraphics[height=6cm,width=8cm]{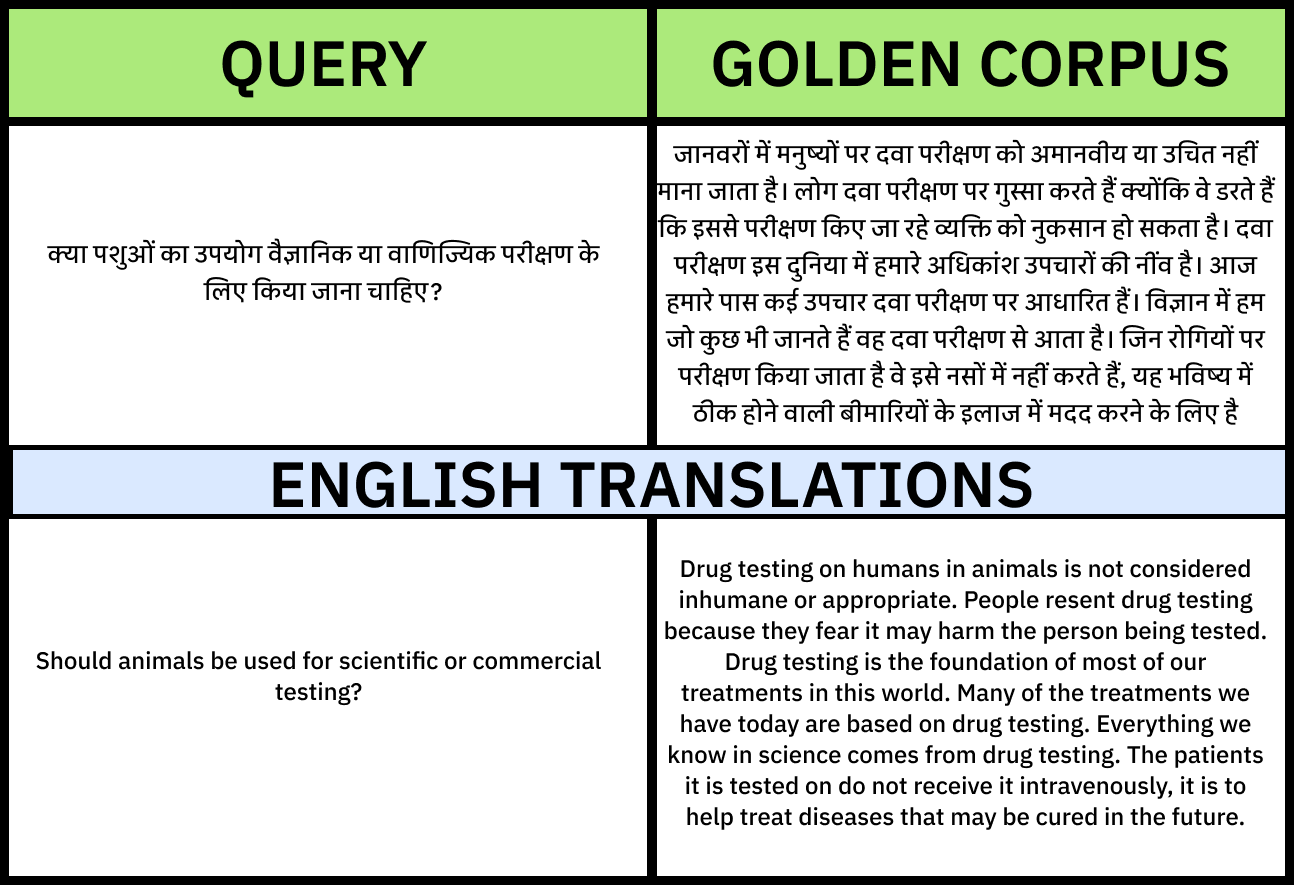}
    \caption{An example of a query with its corresponding golden corpus from the Touch\'e-2020 Dataset}
    \label{fig:touche_example}
\end{figure}

An example of query with its corresponding golden corpus from the Touch\'e-2020 dataset has been provided in Figure \ref{fig:touche_example}

\begin{figure}[hbt!]
    \centering
    \includegraphics[height=4cm,width=8cm]{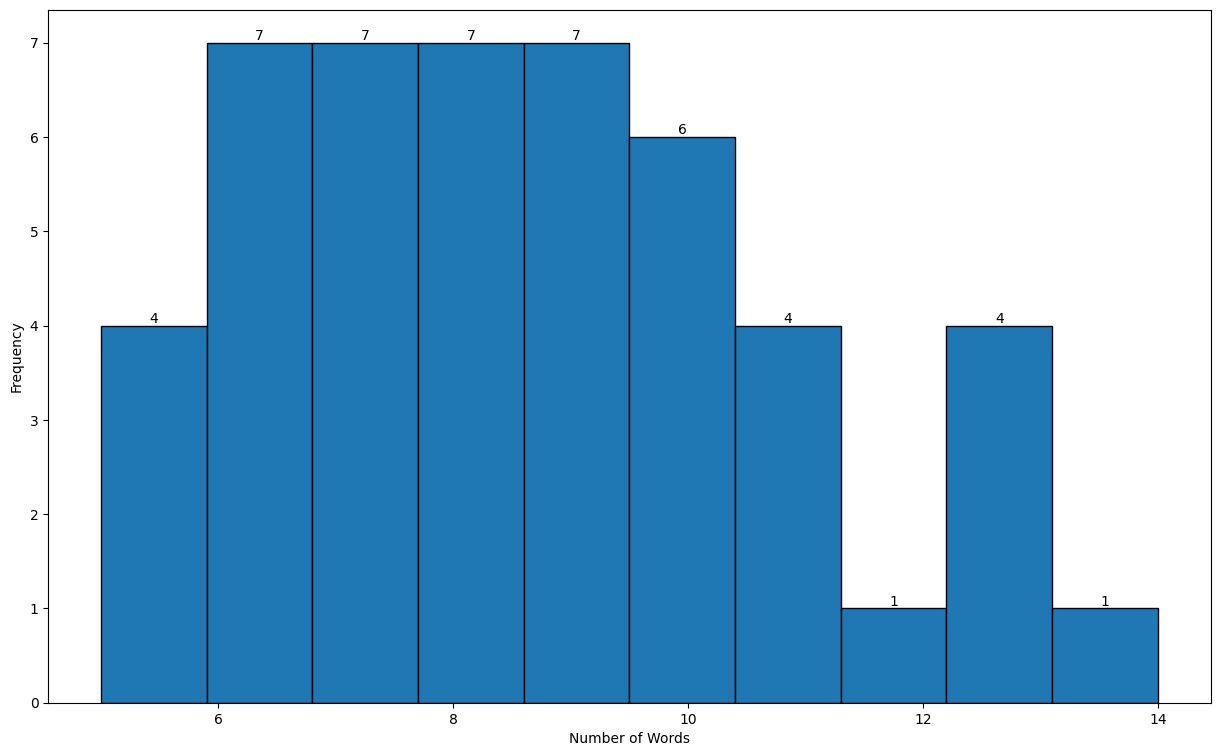}
    \caption{Distribution of the number of words in the queries of Touch\'e-2020 Dataset}
    \label{fig:touche_queries}
\end{figure}

\begin{figure}[hbt!]
    \centering
    \includegraphics[height=4cm,width=8cm]{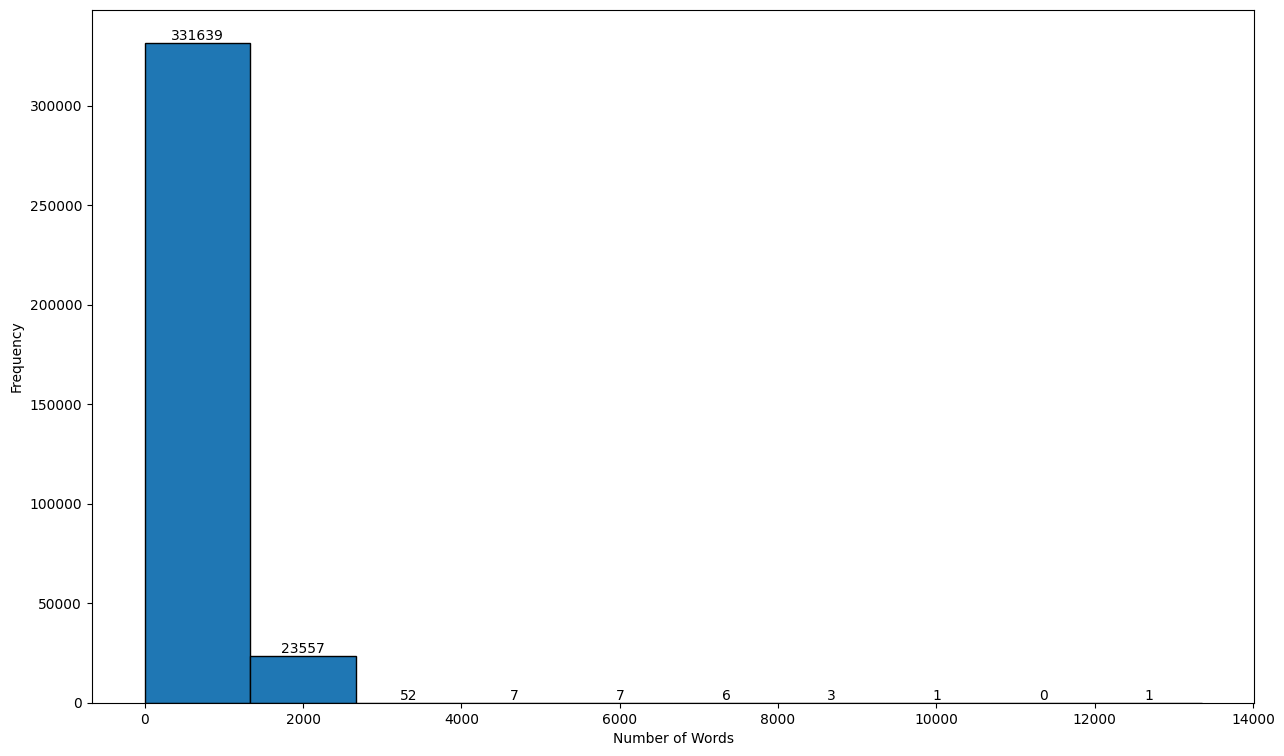}
    \caption{Distribution of Number of Words in corpus of Touch\'e-2020 Dataset}
    \label{fig:touche_corpus}
\end{figure}

Distribution of the number of words in the corpus and queries in Touch\'e-2020 dataset has been shown in Figure \ref{fig:touche_corpus} and Figure \ref{fig:touche_queries} respectively.

\subsubsection{NQ}
\begin{enumerate}
\item \textbf{Task Definition: } The task here is to retrieve the correct answer to a given question from a broad corpus of candidate answers. The NQ datset in the Hindi-BEIR Benchmark is the translated and filtered version of the NQ dataset released by \citeauthor{thakur2021beir} which contains Google search queries and documents with paragraphs and answer spans within Wikipedia articles as the corpus.

    \item \textbf{Domain :} WikiPedia
\end{enumerate} 

\begin{figure}[hbt!]
    \centering
    \includegraphics[height=6cm,width=8cm]{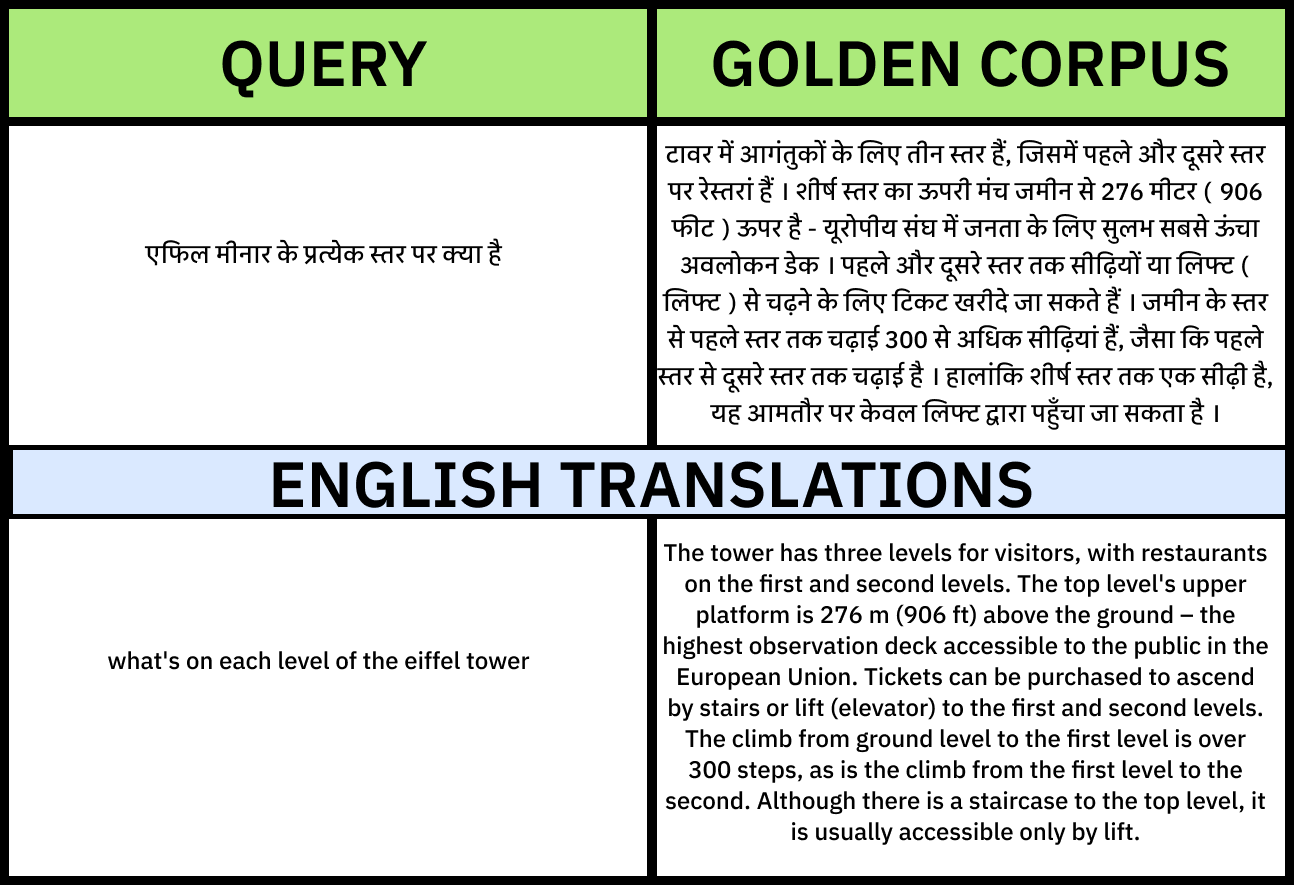}
    \caption{An example of a query with its corresponding golden corpus from the NQ Dataset}
    \label{fig:nq_example}
\end{figure}

An example of query with its corresponding golden corpus from the NQ dataset has been provided in Figure \ref{fig:nq_example}

\begin{figure}[hbt!]
    \centering
    \includegraphics[height=4cm,width=8cm]{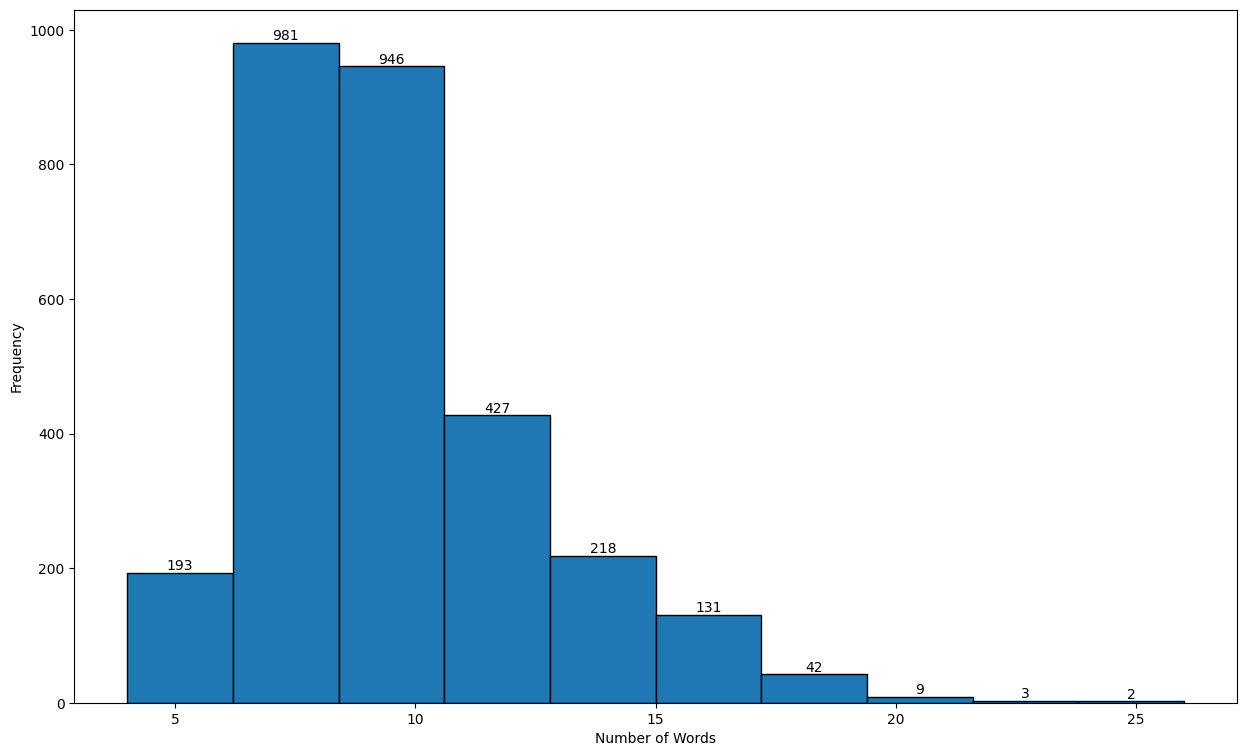}
    \caption{Distribution of the number of words in the queries of NQ Dataset}
    \label{fig:nq_queries}
\end{figure}

\begin{figure}[hbt!]
    \centering
    \includegraphics[height=4cm,width=8cm]{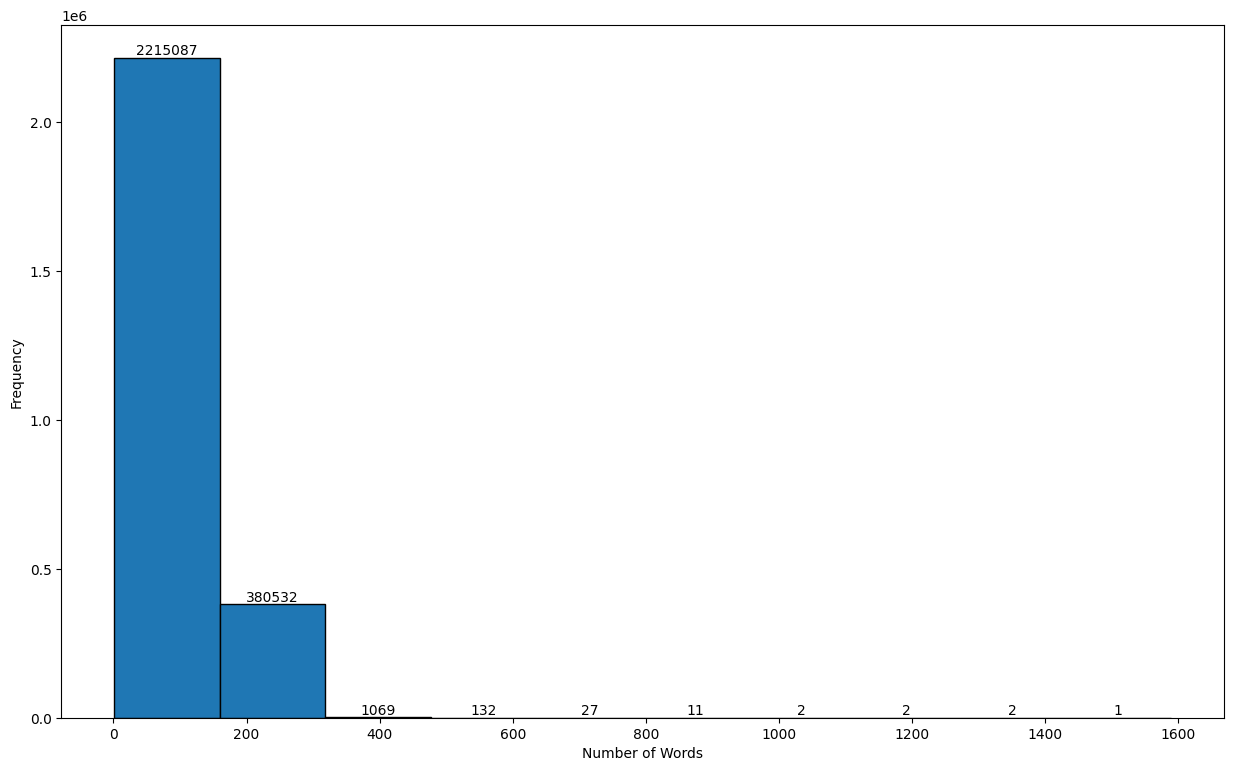}
    \caption{Distribution of Number of Words in corpus of NQ Dataset}
    \label{fig:nq_corpus}
\end{figure}

Distribution of the number of words in the corpus and queries in NQ dataset has been shown in Figure \ref{fig:nq_corpus} and Figure \ref{fig:nq_queries} respectively.

\subsubsection{FEVER}
\begin{figure}[hbt!]
    \centering
    \includegraphics[height=6cm,width=8cm]{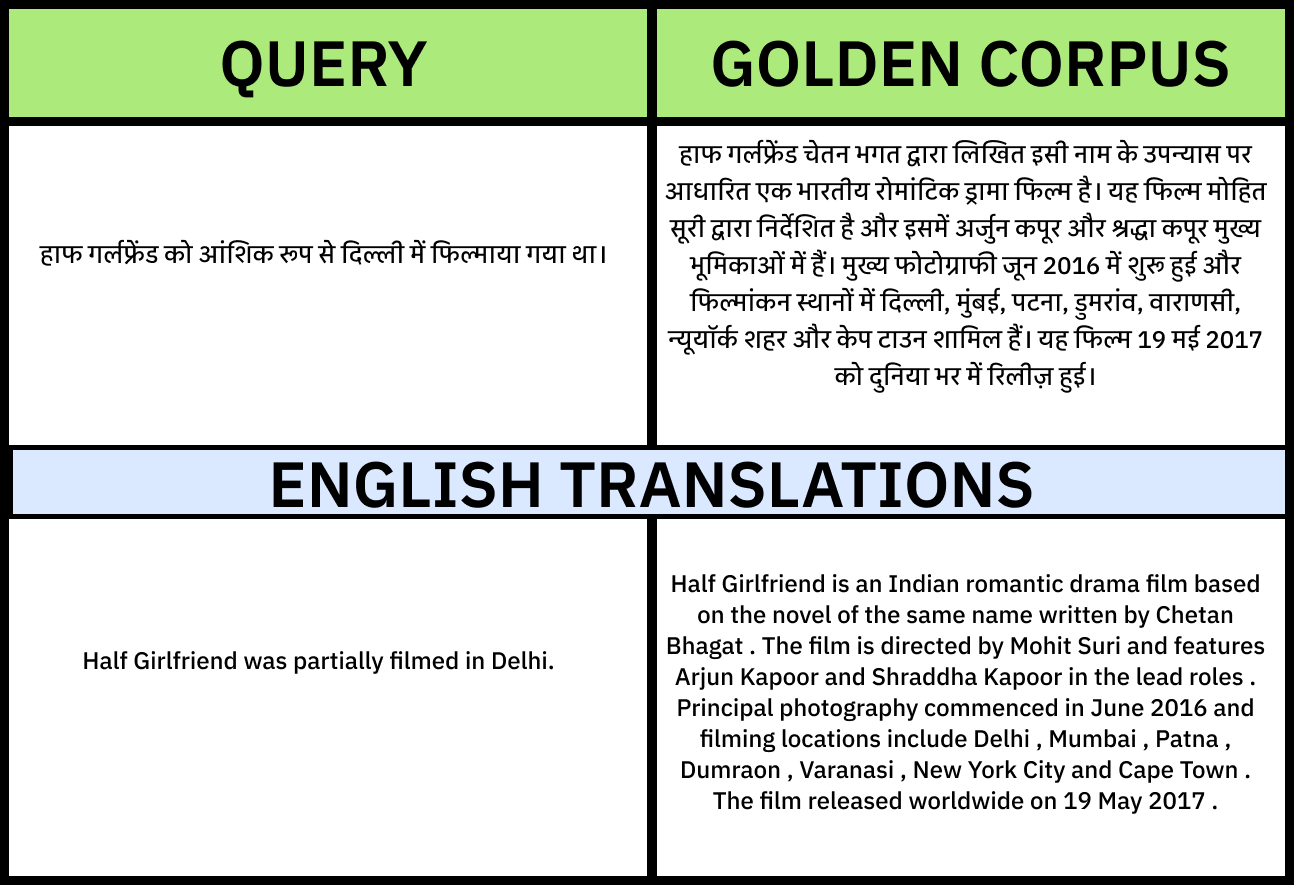}
    \caption{An example of a query with its corresponding golden corpus from the FEVER Dataset}
    \label{fig:fever_example}
\end{figure}

\begin{enumerate}

\item \textbf{Task Definition: } Similar to SciFact, here , the task is to retrieve relevant documents that claim a given fact (query). We translate and filter the test split of the FEVER dataset as proposed by \citeauthor{thakur2021beir} and include it in the Hindi-BEIR Benchmark.
    \item \textbf{Domain :} WikiPedia
\end{enumerate}
An example of query with its corresponding golden corpus from the FEVER dataset has been provided in Figure \ref{fig:fever_example}

\begin{figure}[hbt!]
    \centering
    \includegraphics[height=4cm,width=8cm]{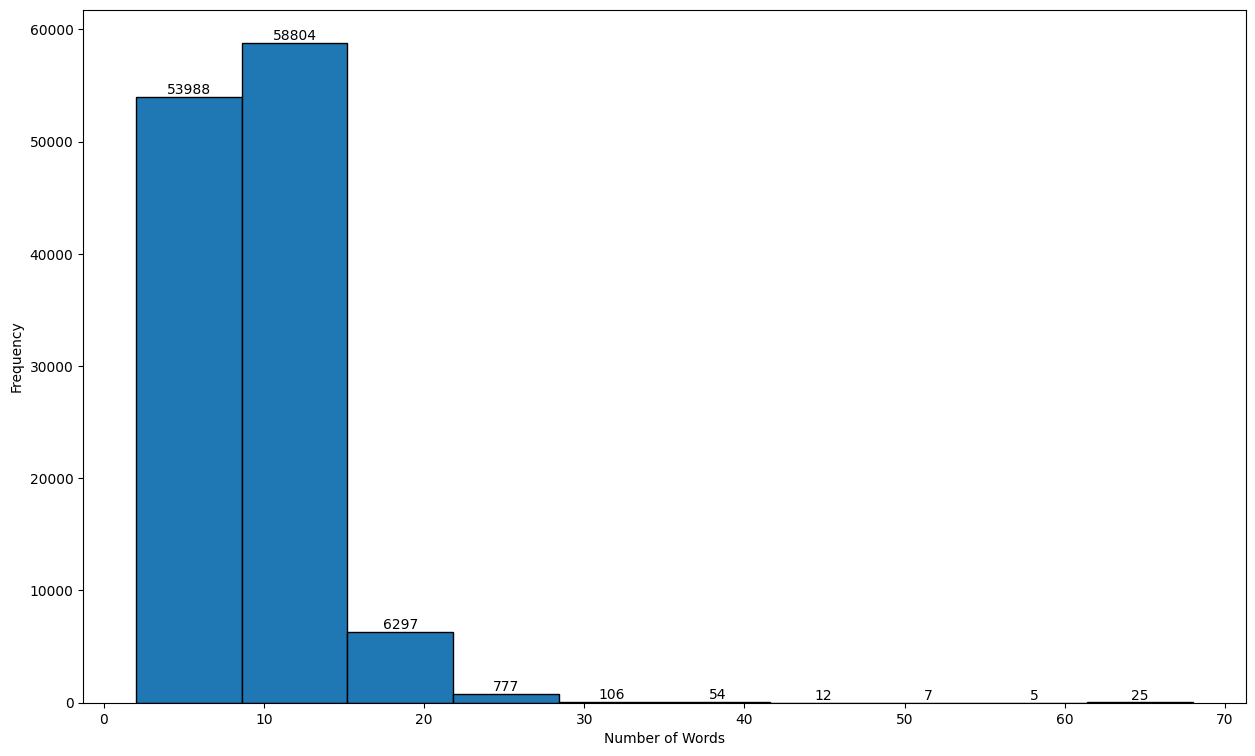}
    \caption{Distribution of the number of words in the queries of FEVER Dataset}
    \label{fig:fever_queries}
\end{figure}

\begin{figure}[hbt!]
    \centering
    \includegraphics[height=4cm,width=8cm]{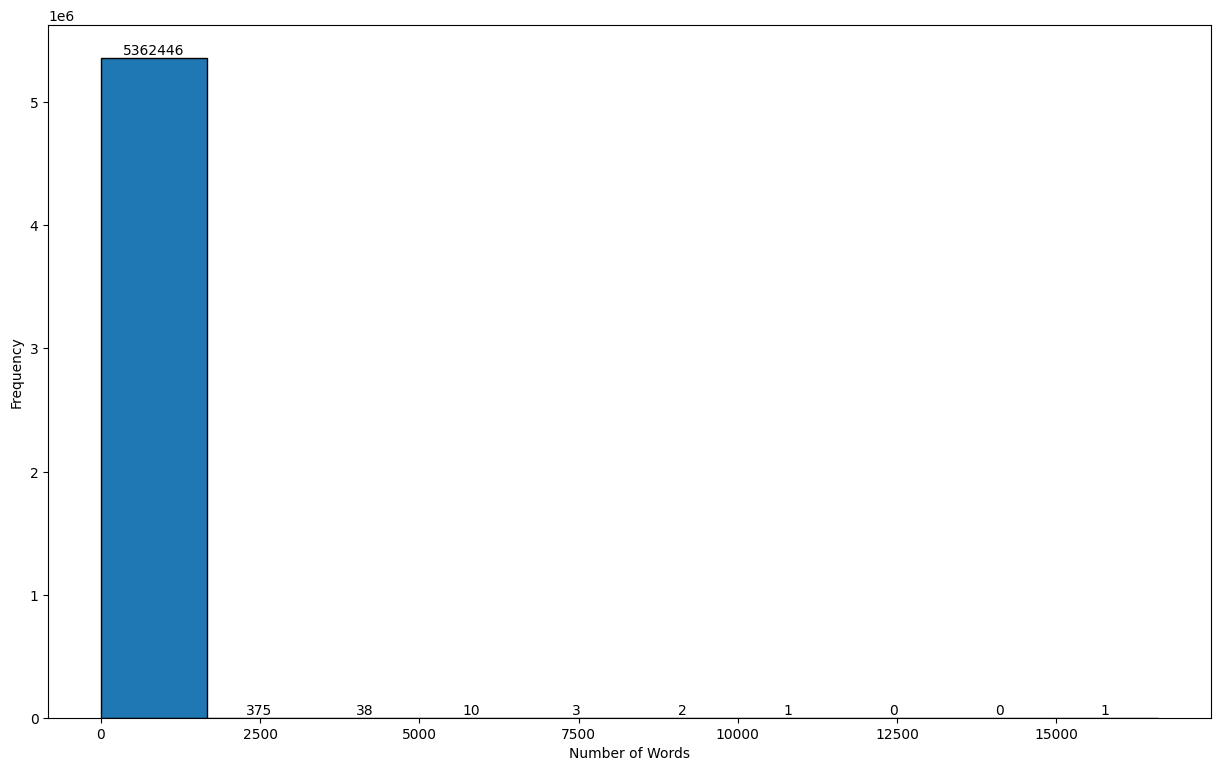}
    \caption{Distribution of Number of Words in corpus of FEVER Dataset}
    \label{fig:fever_corpus}
\end{figure}

Distribution of the number of words in the corpus and queries in FEVER dataset has been shown in Figure \ref{fig:fever_corpus} and Figure \ref{fig:fever_queries} respectively.

\subsubsection{Climate-FEVER}
\begin{figure}[hbt!]
    \centering
    \includegraphics[height=8cm,width=8cm]{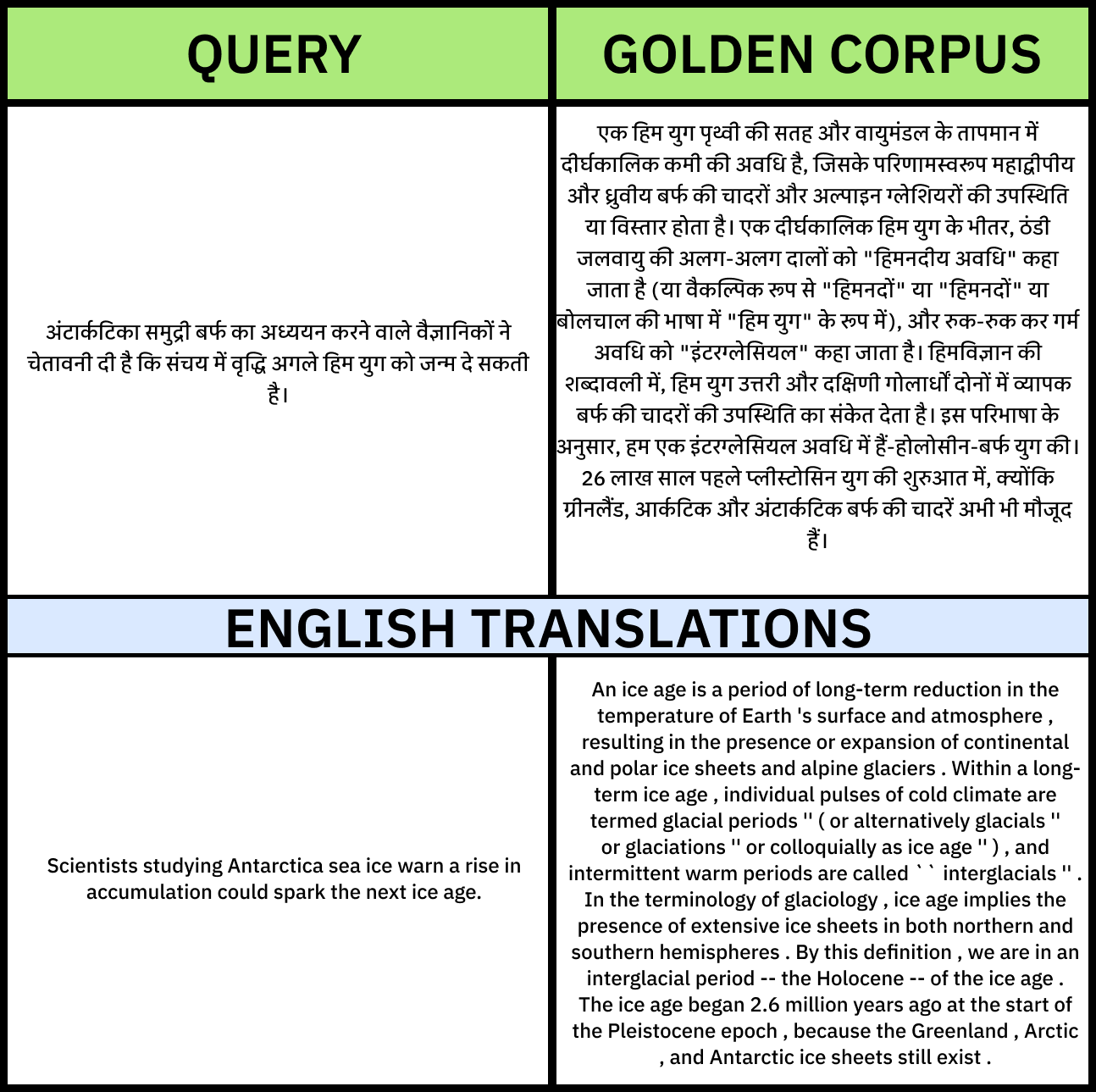}
    \caption{An example of a query with its corresponding golden corpus from the Climate-FEVER Dataset}
    \label{fig:climate_fever_example}
\end{figure}

\begin{enumerate}

\item \textbf{Task Definition: } Similar to FEVER, Climate-FEVER is a dataset for verification of real-world climate claims. We translate and filter the test split of the Climate-FEVER dataset as proposed by \citeauthor{thakur2021beir} and include it in the Hindi-BEIR Benchmark.
    \item \textbf{Domain :} WikiPedia
\end{enumerate}
An example of query with its corresponding golden corpus from the Climate-FEVER dataset has been provided in Figure \ref{fig:climate_fever_example}

\begin{figure}[hbt!]
    \centering
    \includegraphics[height=4cm,width=8cm]{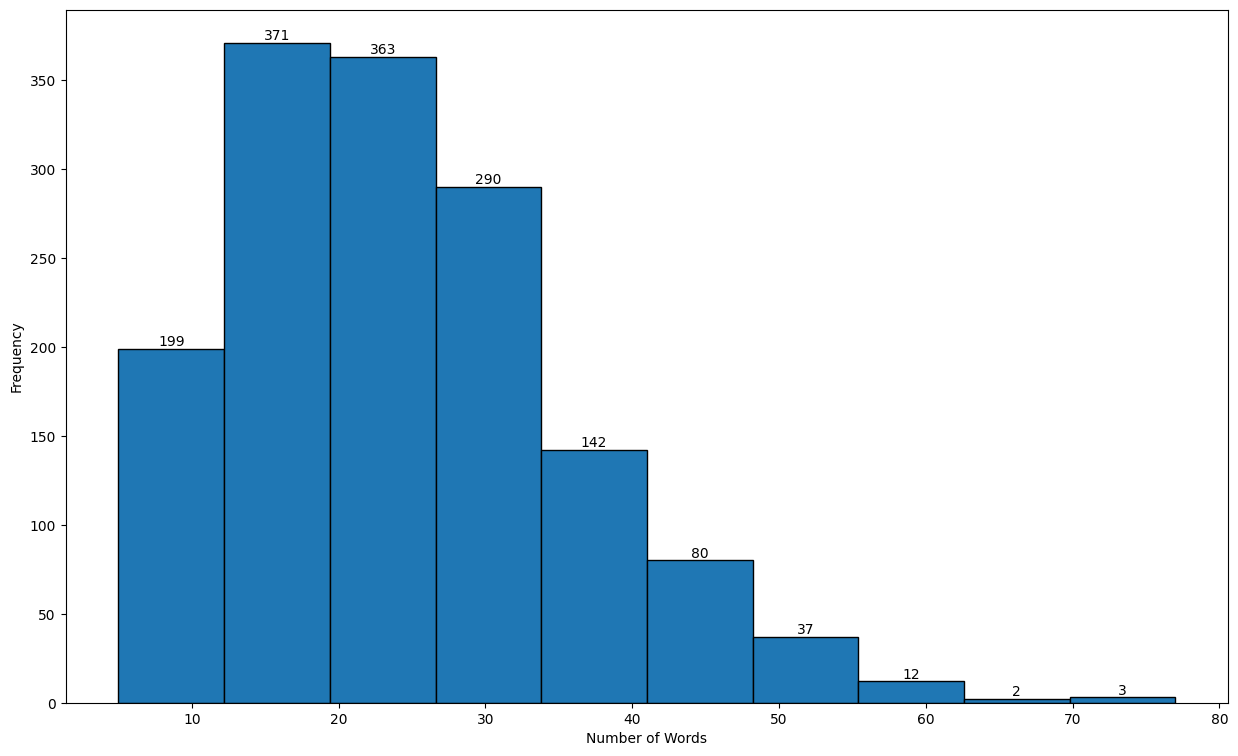}
    \caption{Distribution of the number of words in the queries of Climate-FEVER Dataset}
    \label{fig:climate_fever_queries}
\end{figure}

\begin{figure}[hbt!]
    \centering
    \includegraphics[height=4cm,width=8cm]{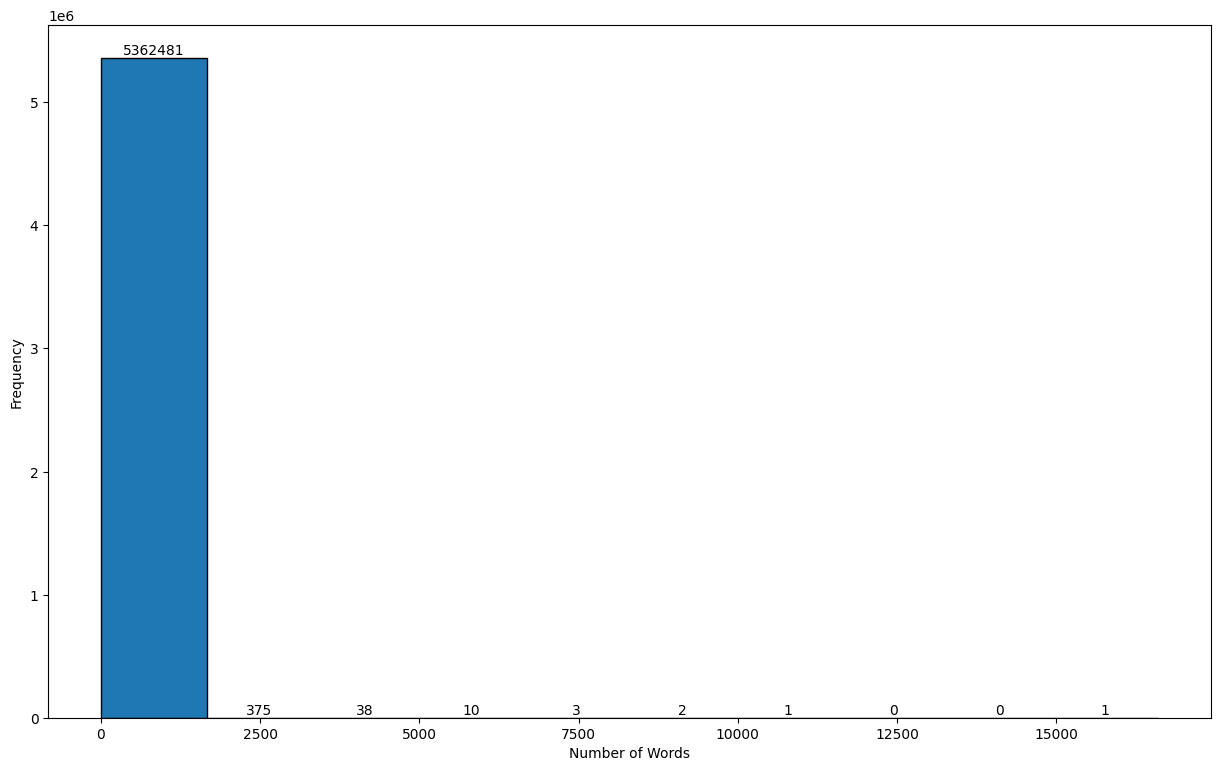}
    \caption{Distribution of Number of Words in corpus of Climate-FEVER Dataset}
    \label{fig:climate_fever_corpus}
\end{figure}

Distribution of the number of words in the corpus and queries in Climate-FEVER dataset has been shown in Figure \ref{fig:climate_fever_corpus} and Figure \ref{fig:climate_fever_queries} respectively.

\subsubsection{CC News Retrieval}

\begin{figure}[hbt!]
    \centering
    \includegraphics[height=8.5cm,width=8cm]{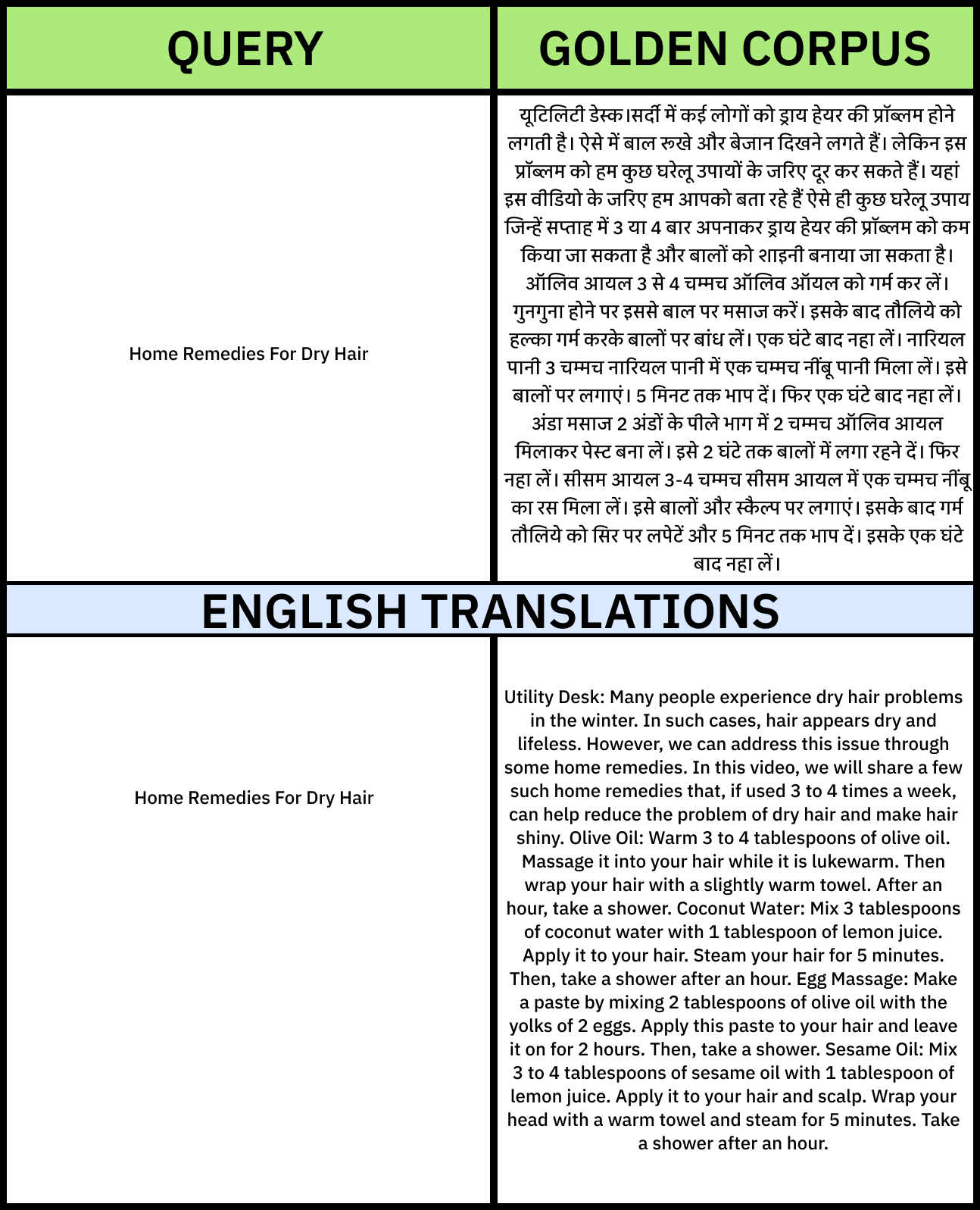}
    \caption{An example of a query with its corresponding golden corpus from the CC News Retrieval Dataset}
    \label{fig:ccnews_example}
\end{figure}

\begin{enumerate}

\item \textbf{Task Definition: } CC News Retrieval Introduces the task to retrieving relevant news articles given a news title. Subsection \ref{cc_news} talks in details about the data creation process.
    \item \textbf{Domain :} News
\end{enumerate}
An example of query with its corresponding golden corpus from the CC News Retrieval dataset has been provided in Figure \ref{fig:ccnews_example}

\begin{figure}[hbt!]
    \centering
    \includegraphics[height=4cm,width=8cm]{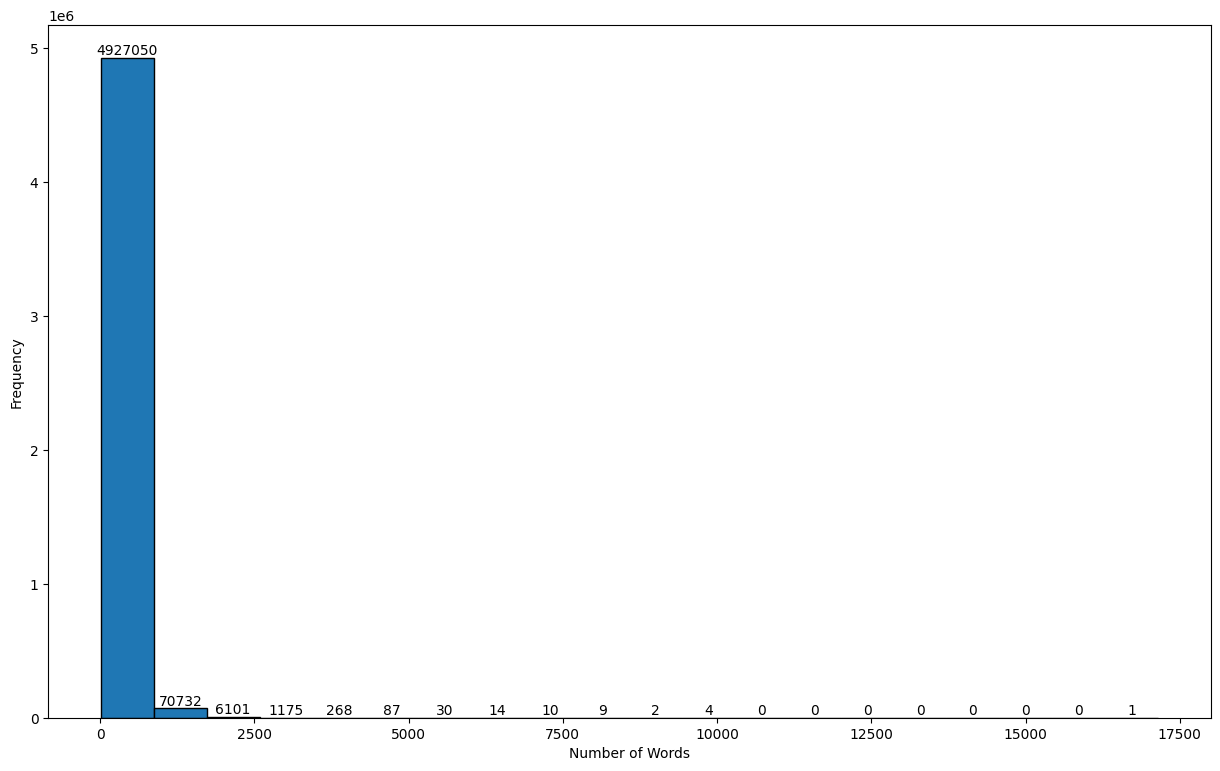}
    \caption{Distribution of Number of Words in corpus of CC News Retrieval Dataset}
    \label{fig:ccnews_corpus}
\end{figure}

Distribution of the number of words in the corpus and queries in CC News Retrieval dataset has been shown in Figure \ref{fig:ccnews_corpus} and Figure \ref{fig:climate_fever_queries} respectively.
\begin{figure}[hbt!]
    \centering
    \includegraphics[height=4cm,width=8cm]{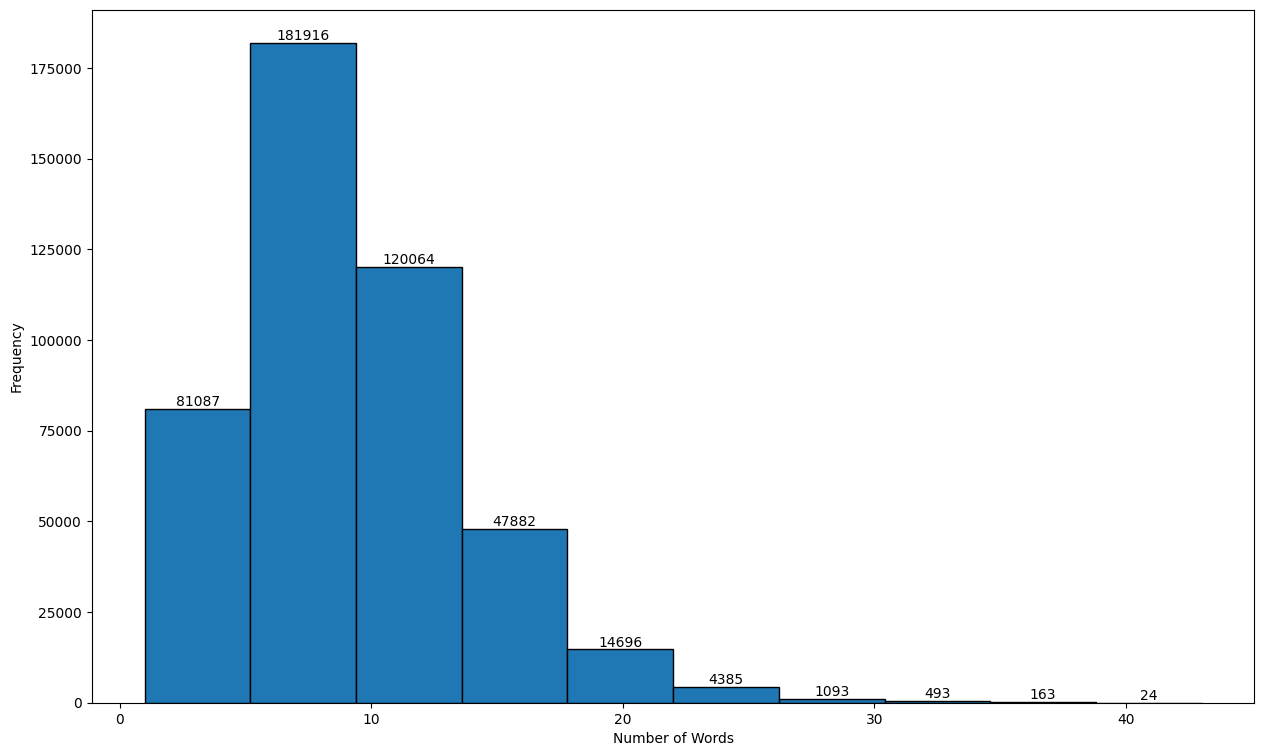}
    \caption{Distribution of the number of words in the queries of CC News Retrieval Dataset}
    \label{fig:ccnews_queries}
\end{figure}

\subsubsection{MLDR}
\begin{figure}[hbt!]
    \centering
    \includegraphics[height=8cm,width=8cm]{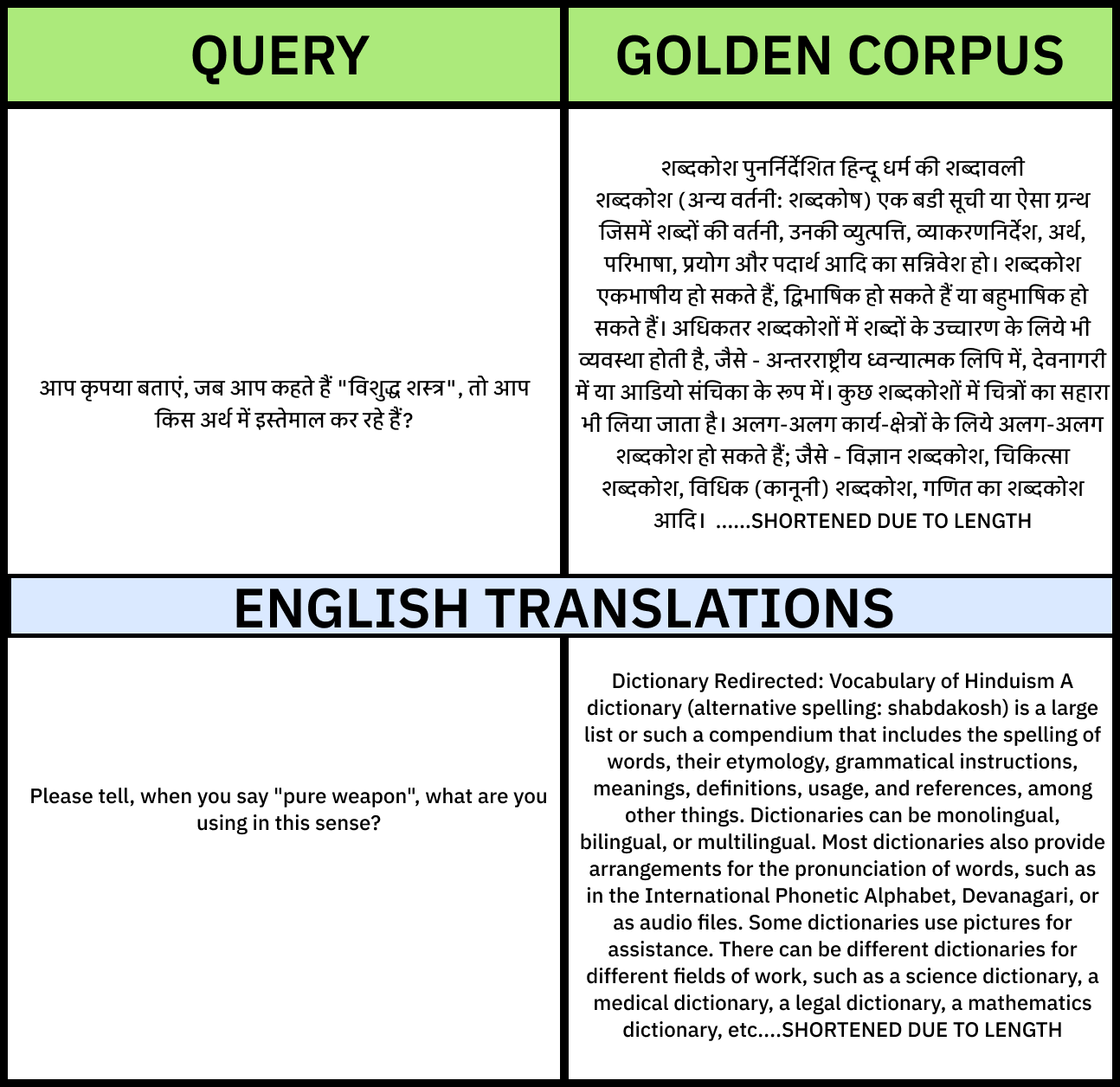}
    \caption{An example of a query with its corresponding golden corpus from the MLDR Dataset}
    \label{fig:mldr_example}
\end{figure}

\begin{enumerate}

\item \textbf{Task Definition: } Introduced by \citeauthor{chen2024bge} : MLDR (Multilingual Long-Doc Retrieval) is curated by the multilingual articles from Wikipedia, Wudao and mC4. These consists of long corpus some of which contains more than 10000 words.
    \item \textbf{Domain :} Miscellaneous
\end{enumerate}
An example of query with its corresponding golden corpus from the MLDR dataset has been provided in Figure \ref{fig:mldr_example}

\begin{figure}[hbt!]
    \centering
    \includegraphics[height=4cm,width=8cm]{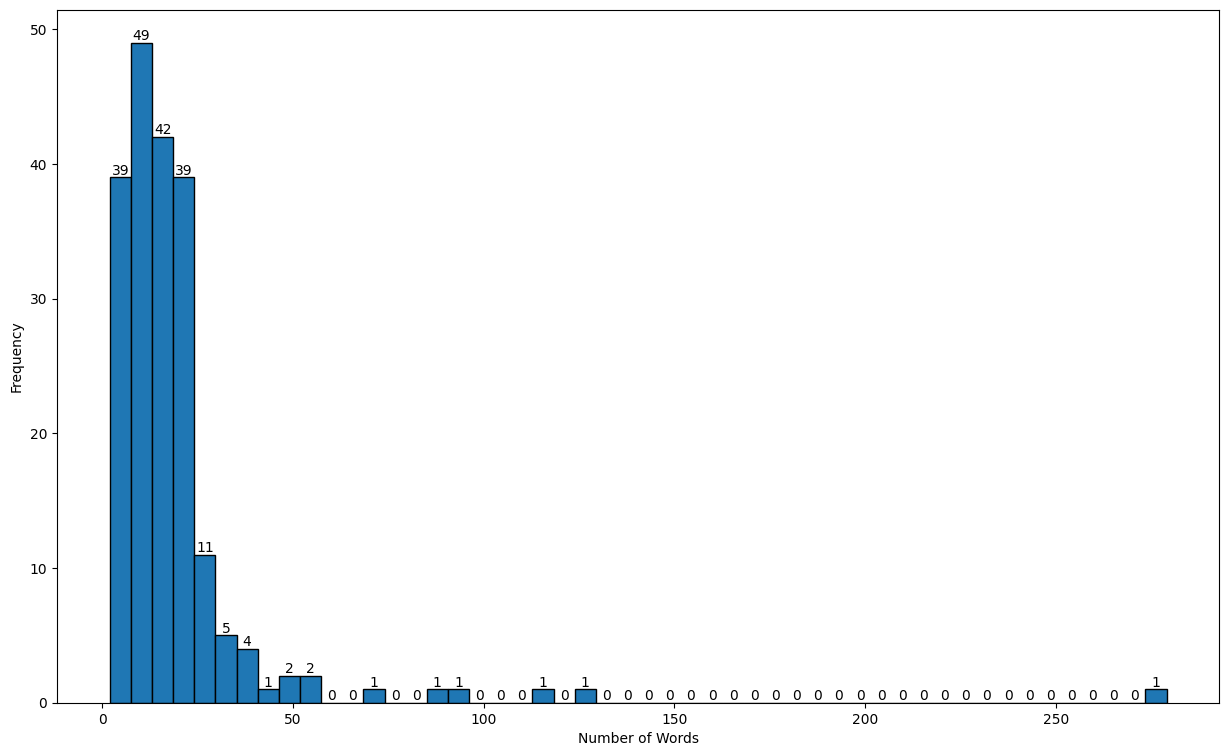}
    \caption{Distribution of the number of words in the queries of MLDR Dataset}
    \label{fig:mldr_queries}
\end{figure}

\begin{figure}[hbt!]
    \centering
    \includegraphics[height=4cm,width=8cm]{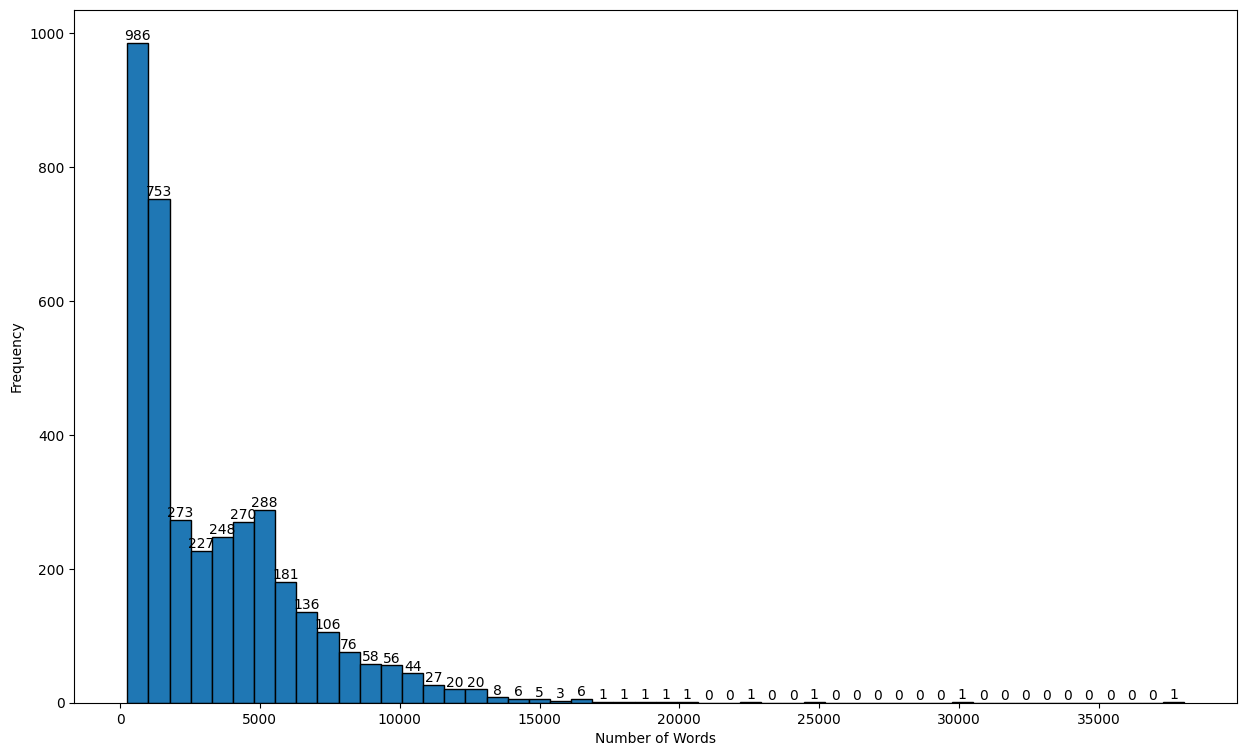}
    \caption{Distribution of Number of Words in corpus of MLDR Dataset}
    \label{fig:mldr_corpus}
\end{figure}

Distribution of the number of words in the corpus and queries in MLDR dataset has been shown in Figure \ref{fig:mldr_corpus} and Figure \ref{fig:mldr_queries} respectively.

\subsubsection{MIRACL}

\begin{figure}[hbt!]
    \centering
    \includegraphics[height=8cm,width=8cm]{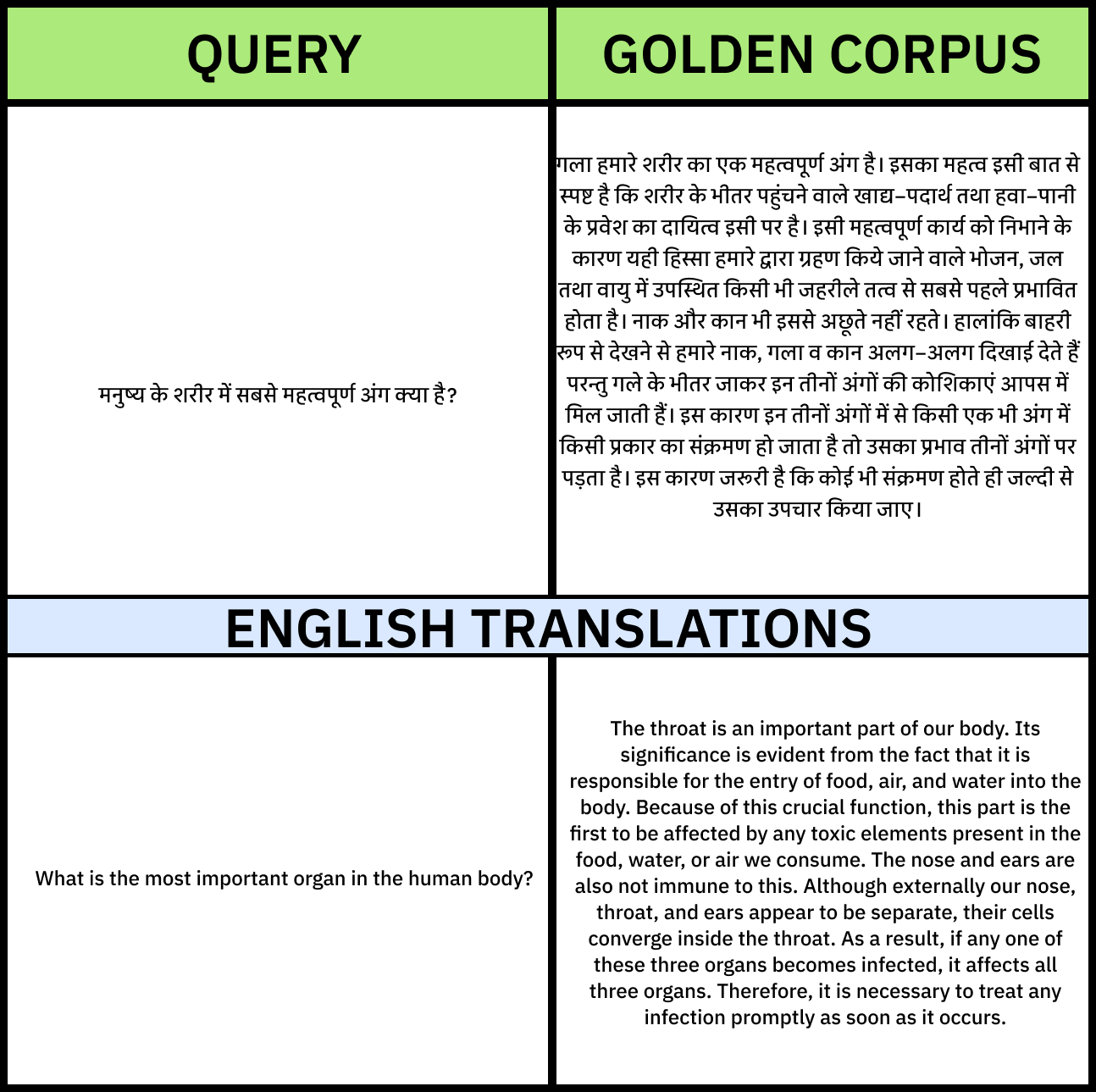}
    \caption{An example of a query with its corresponding golden corpus from the MIRACL Dataset}
    \label{fig:miracl_example}
\end{figure}

\begin{enumerate}

\item \textbf{Task Definition: } MIRACL has been created from the wikipedia dump of in each language. Only the plain text is considered while images etc. are omited. The plain text is split up into multiple paragraphs which act as the corpus. We consider the dev split of Hindi subset of the original MIRACL in the Hindi-BEIR Becnhmark.

    \item \textbf{Domain :} WikiPedia
\end{enumerate}
An example of query with its corresponding golden corpus from the MIRACL dataset has been provided in Figure \ref{fig:miracl_example}

\begin{figure}[hbt!]
    \centering
    \includegraphics[height=4cm,width=8cm]{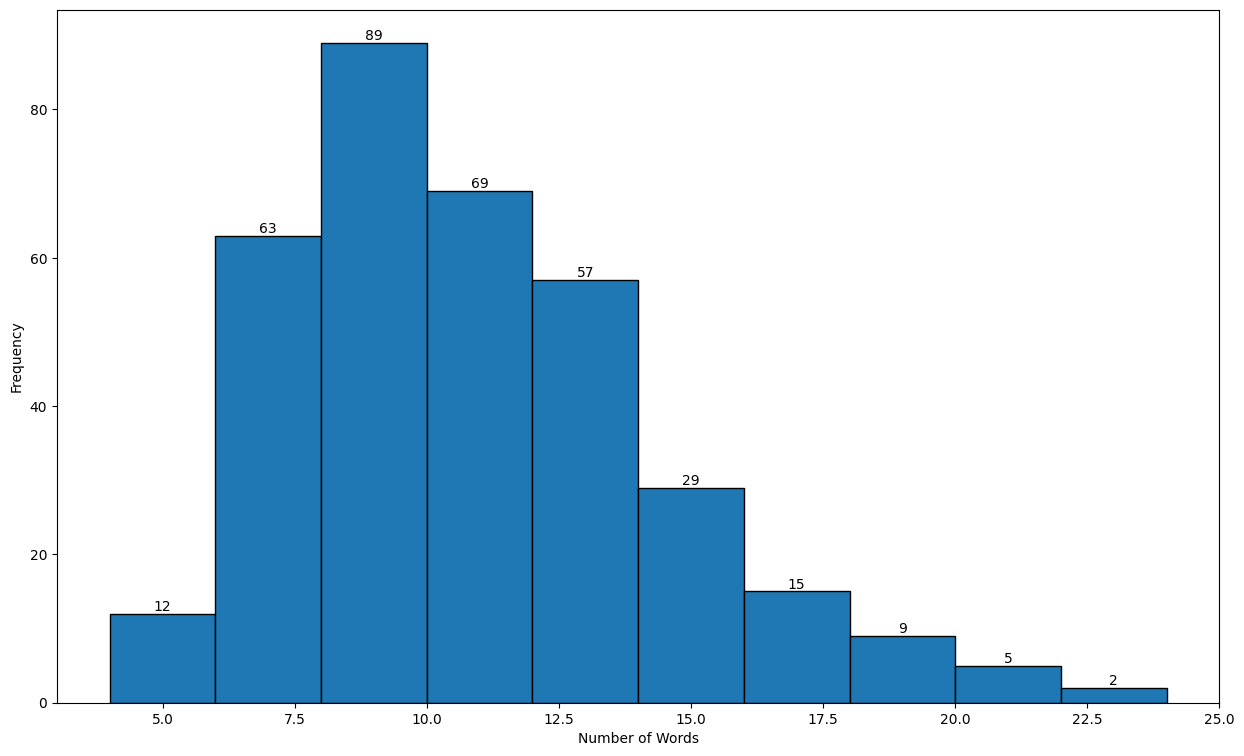}
    \caption{Distribution of the number of words in the queries of MIRACL Dataset}
    \label{fig:miracl_queries}
\end{figure}

\begin{figure}[hbt!]
    \centering
    \includegraphics[height=4cm,width=8cm]{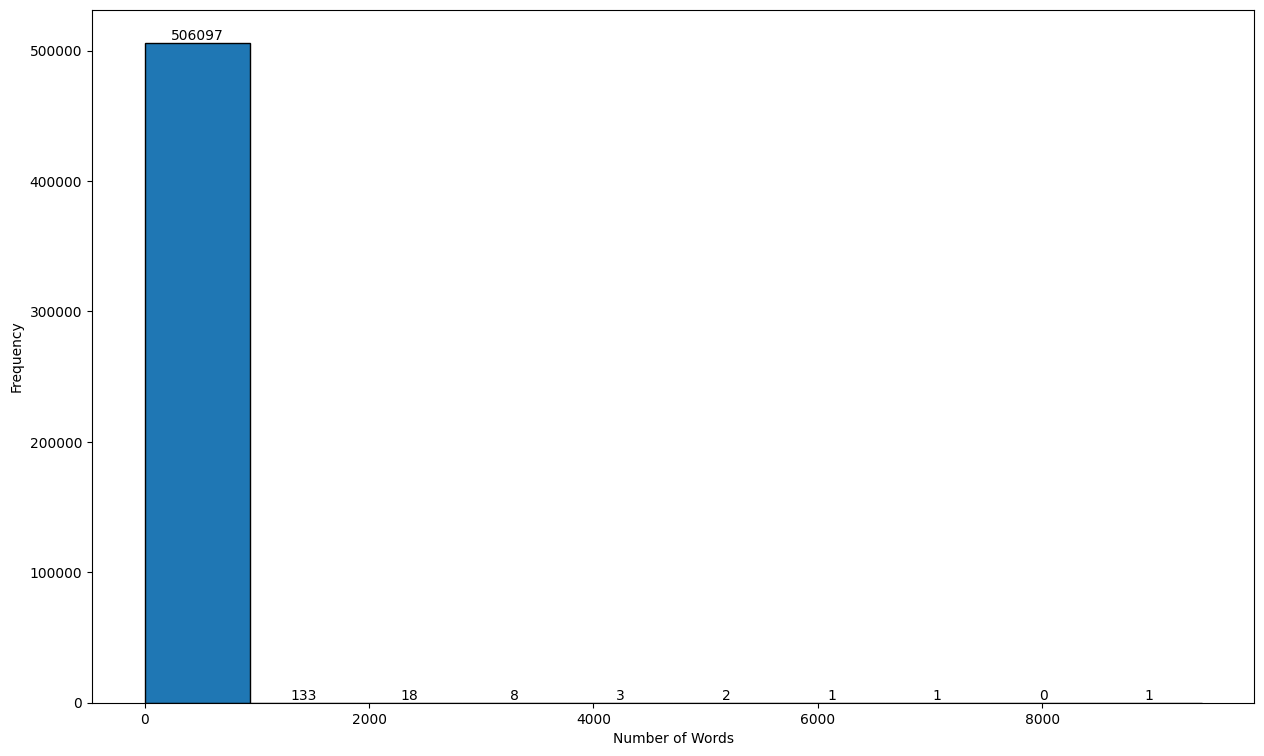}
    \caption{Distribution of Number of Words in corpus of MIRACL Dataset}
    \label{fig:miracl_corpus}
\end{figure}

Distribution of the number of words in the corpus and queries in MIRACL dataset has been shown in Figure \ref{fig:miracl_corpus} and Figure \ref{fig:miracl_queries} respectively.

\subsubsection{IndicQARetrieval}
\begin{figure}[hbt!]
    \centering
    \includegraphics[height=8cm,width=8cm]{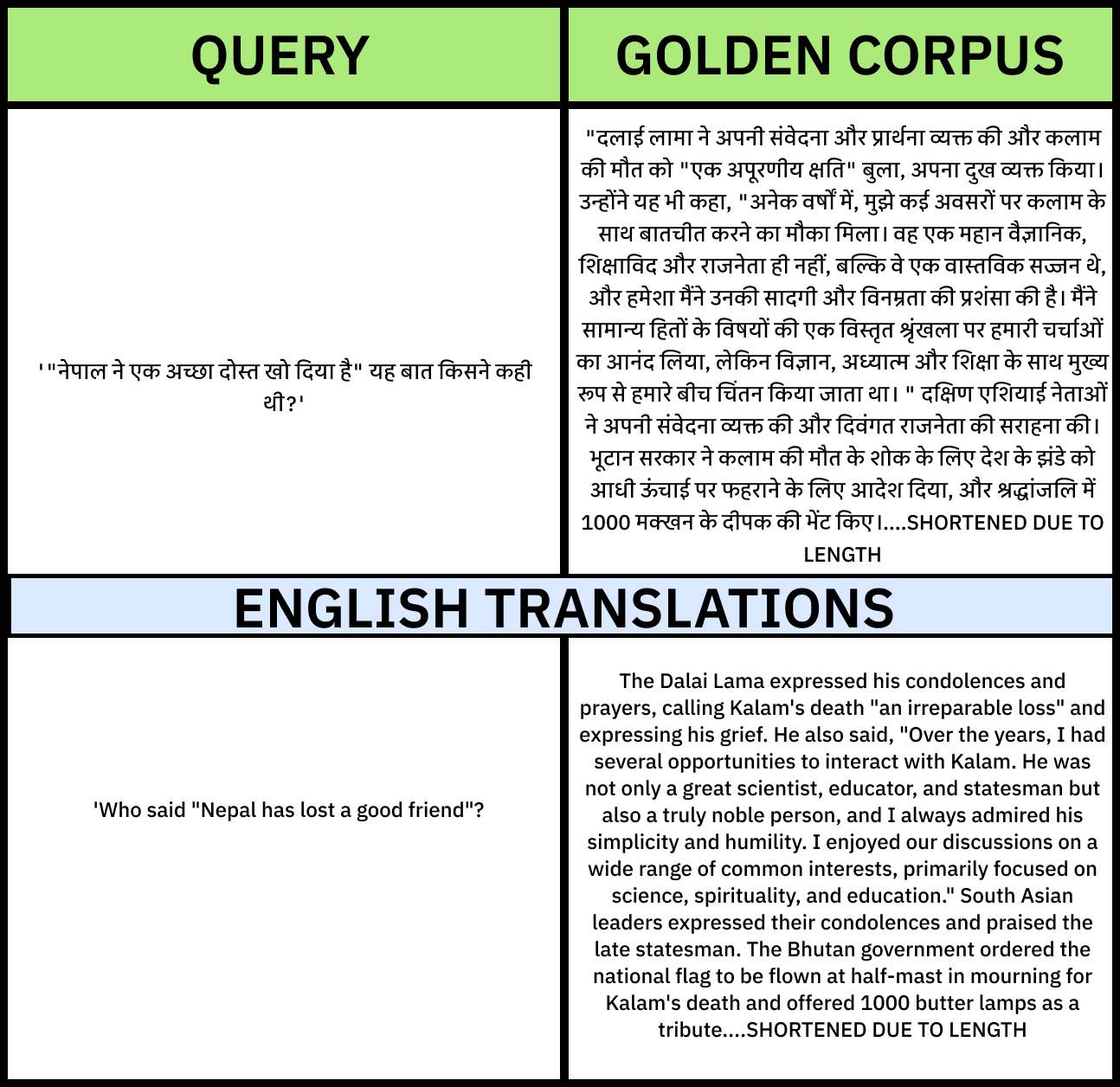}
    \caption{An example of a query with its corresponding golden corpus from the IndicQARetrieval Dataset}
    \label{fig:indicqa_example}
\end{figure}

\begin{enumerate}

\item \textbf{Task Definition: } It is created by transforming the IndicQA dataset \footnote{\url{https://huggingface.co/datasets/ai4bharat/IndicQA}} to a retrieval dataset. The task is similar to NQ, where given a question, the model is expected to return the paragraph which contains the answer to the question. 

    \item \textbf{Domain :} WikiPedia
\end{enumerate}
An example of query with its corresponding golden corpus from the IndicQARetrieval dataset has been provided in Figure \ref{fig:indicqa_example}

\begin{figure}[hbt!]
    \centering
    \includegraphics[height=4cm,width=8cm]{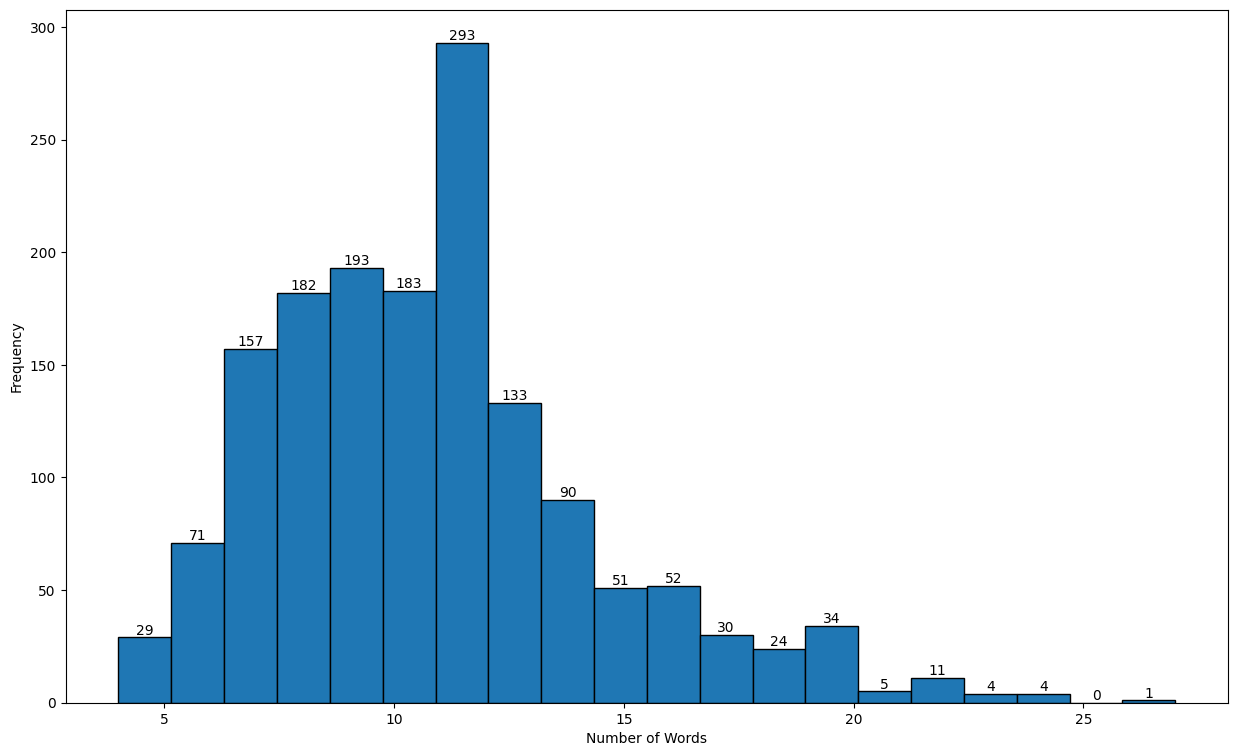}
    \caption{Distribution of the number of words in the queries of IndicQARetrieval Dataset}
    \label{fig:indicqa_queries}
\end{figure}

\begin{figure}[hbt!]
    \centering
    \includegraphics[height=4cm,width=8cm]{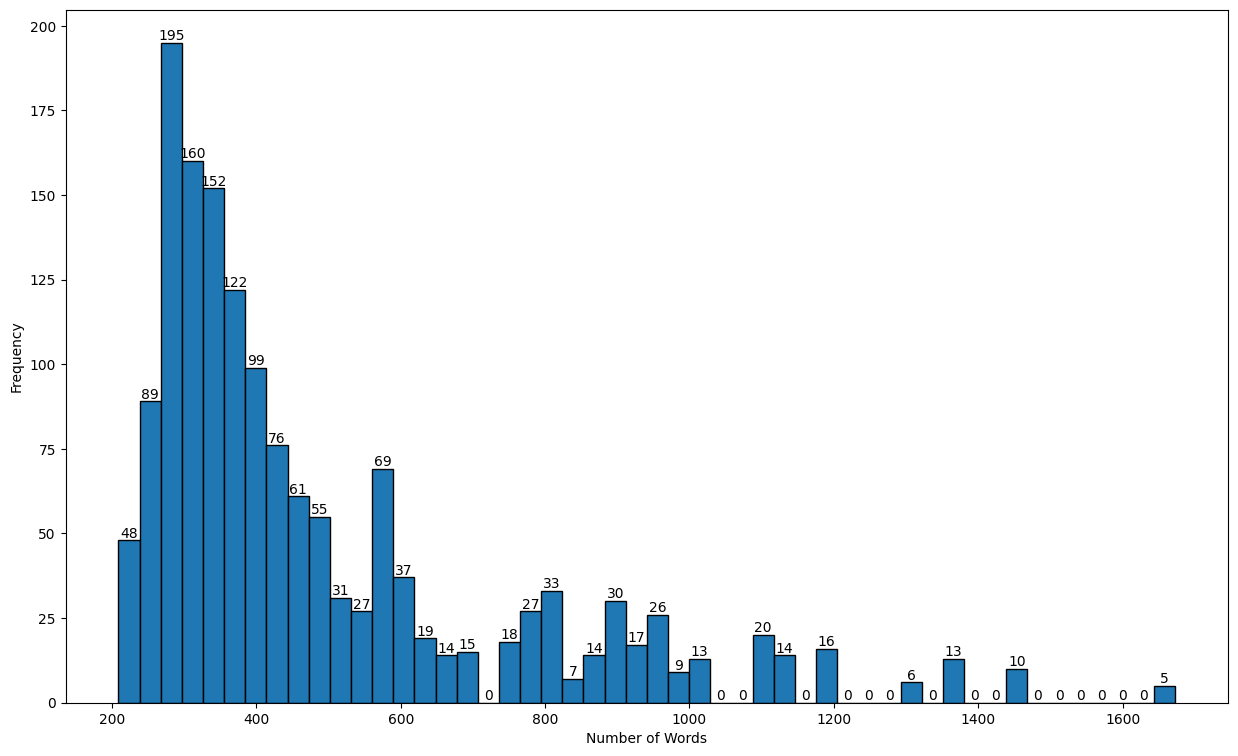}
    \caption{Distribution of Number of Words in corpus of IndicQARetrieval Dataset}
    \label{fig:indicqa_corpus}
\end{figure}

Distribution of the number of words in the corpus and queries in IndicQARetrieval dataset has been shown in Figure \ref{fig:indicqa_corpus} and Figure \ref{fig:indicqa_queries} respectively.

\subsubsection{mMARCO}
\begin{figure}[hbt!]
    \centering
    \includegraphics[height=6cm,width=8cm]{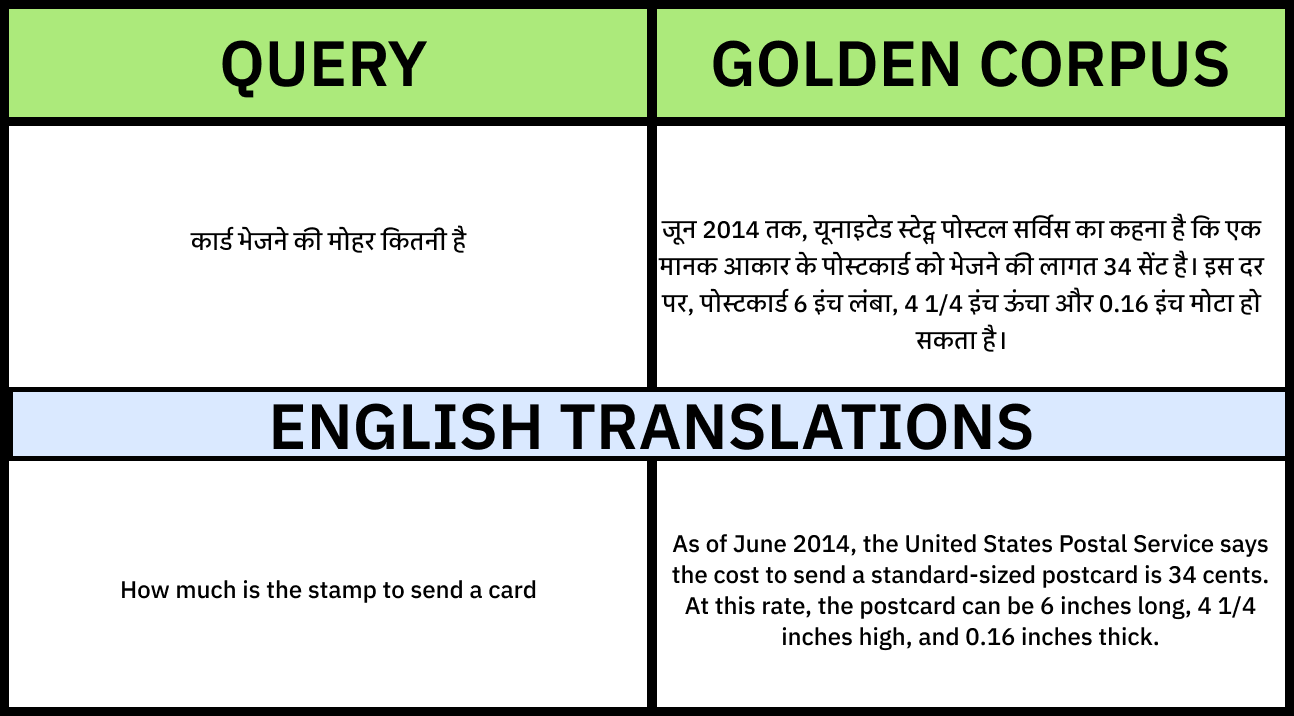}
    \caption{An example of a query with its corresponding golden corpus from the mMARCO Dataset}
    \label{fig:mmarco_example}
\end{figure}

\begin{enumerate}

\item \textbf{Task Definition: } It is a multilingual version of the MSMARCO dataset.The dataset contains translation of queries from Bing search logs with one text passage from various web sources annotated as relevant.
    \item \textbf{Domain :} Miscellaneous
\end{enumerate}
An example of query with its corresponding golden corpus from the mMARCO dataset has been provided in Figure \ref{fig:mmarco_example}

\begin{figure}[hbt!]
    \centering
    \includegraphics[height=4cm,width=8cm]{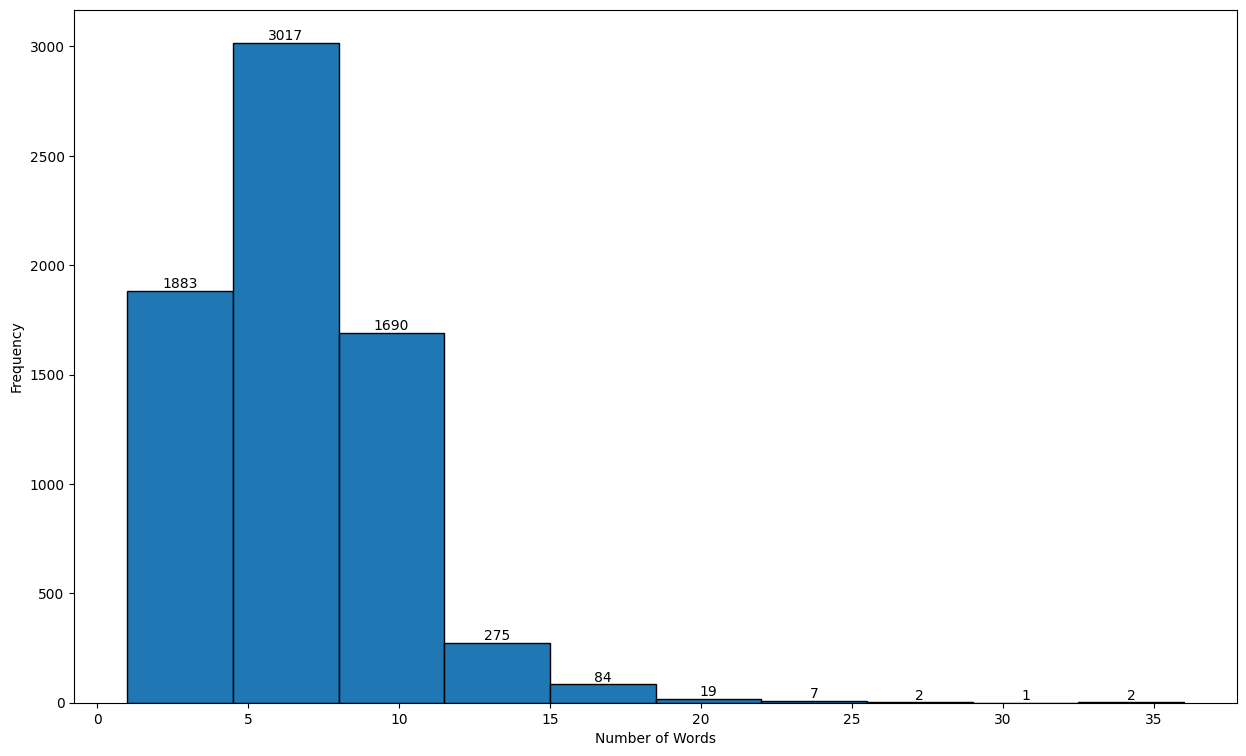}
    \caption{Distribution of the number of words in the queries of mMARCO Dataset}
    \label{fig:mmarco_queries}
\end{figure}

\begin{figure}[hbt!]
    \centering
    \includegraphics[height=4cm,width=8cm]{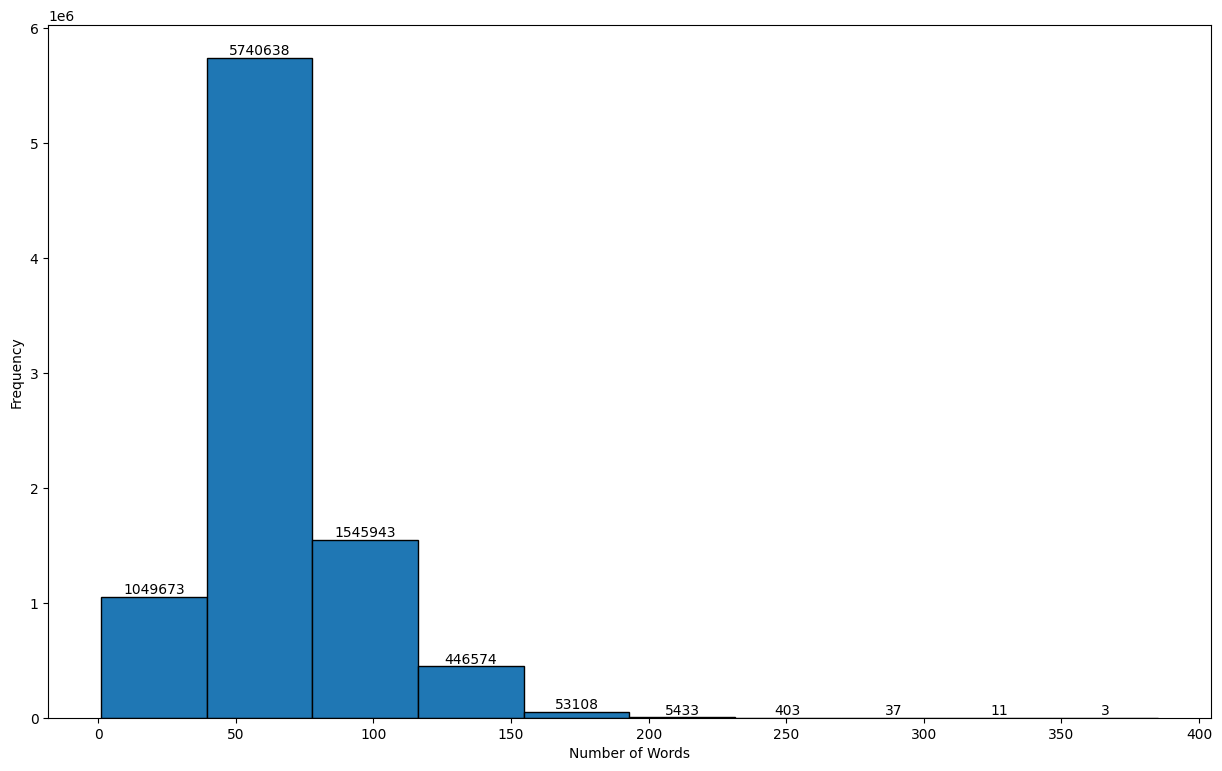}
    \caption{Distribution of Number of Words in corpus of mMARCO Dataset}
    \label{fig:mmarco_corpus}
\end{figure}

Distribution of the number of words in the corpus and queries in mMARCO dataset has been shown in Figure \ref{fig:mmarco_corpus} and Figure \ref{fig:mmarco_queries} respectively.

\subsubsection{WikiPediaRetrieval}
\begin{figure}[hbt!]
    \centering
    \includegraphics[height=6cm,width=8cm]{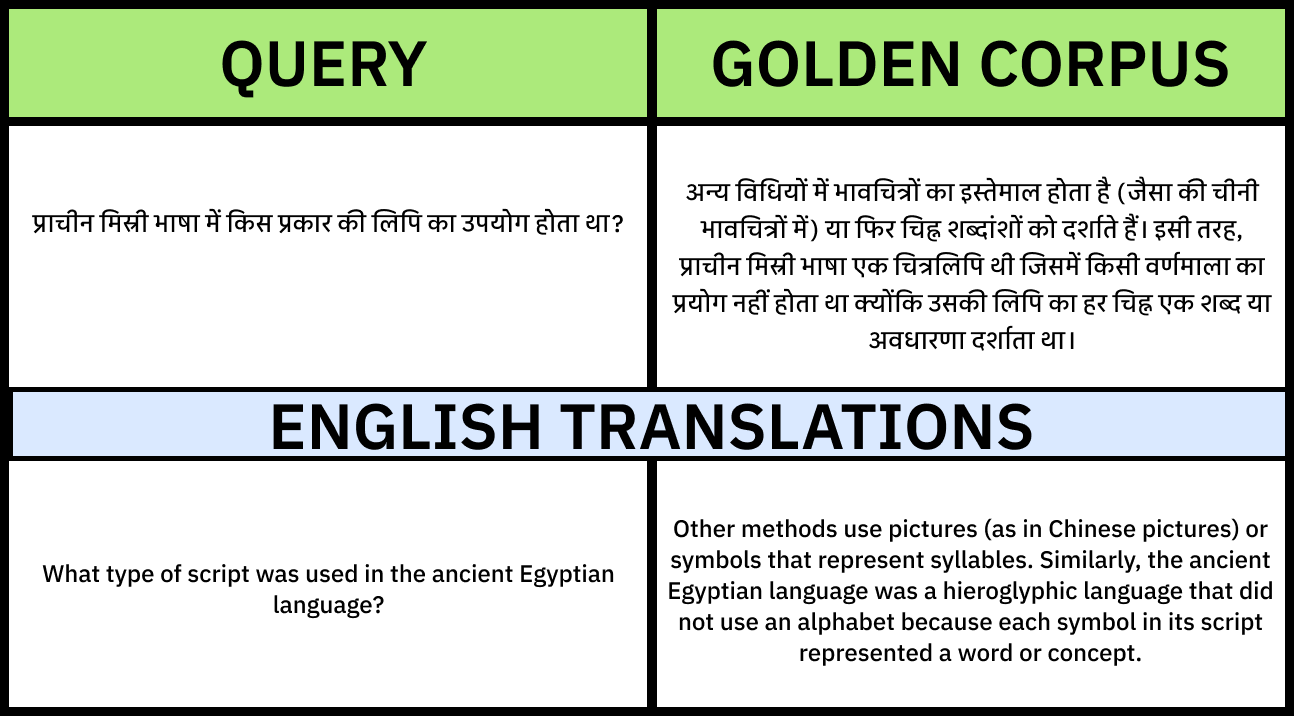}
    \caption{An example of a query with its corresponding golden corpus from the WikiPediaRetrieval Dataset}
    \label{fig:wiki_example}
\end{figure}

\begin{enumerate}

\item \textbf{Task Definition: } It is similar to a question answering task dataset like NQ where given a query, the model is expected to retrieve relevant article which answers the question. We have included the Hindi subset of WikiPediaRetrieval dataset \footnote{\url{https://huggingface.co/collections/ellamind/mmteb-6661723dc229e1da8e837cdf}}

    \item \textbf{Domain :} WikiPedia
\end{enumerate}
An example of query with its corresponding golden corpus from the WikiPediaRetrieval dataset has been provided in Figure \ref{fig:wiki_example}

\begin{figure}[hbt!]
    \centering
    \includegraphics[height=4cm,width=8cm]{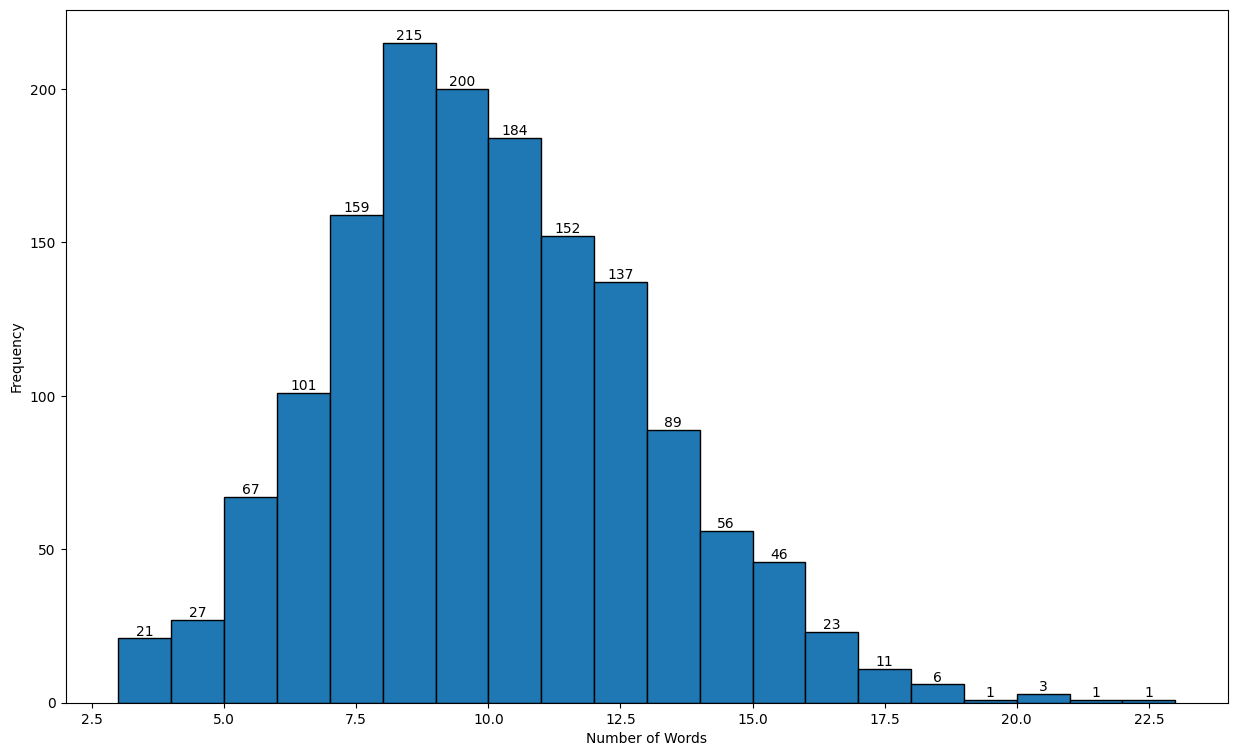}
    \caption{Distribution of the number of words in the queries of WikiPediaRetrieval Dataset}
    \label{fig:wiki_queries}
\end{figure}

\begin{figure}[hbt!]
    \centering
    \includegraphics[height=4cm,width=8cm]{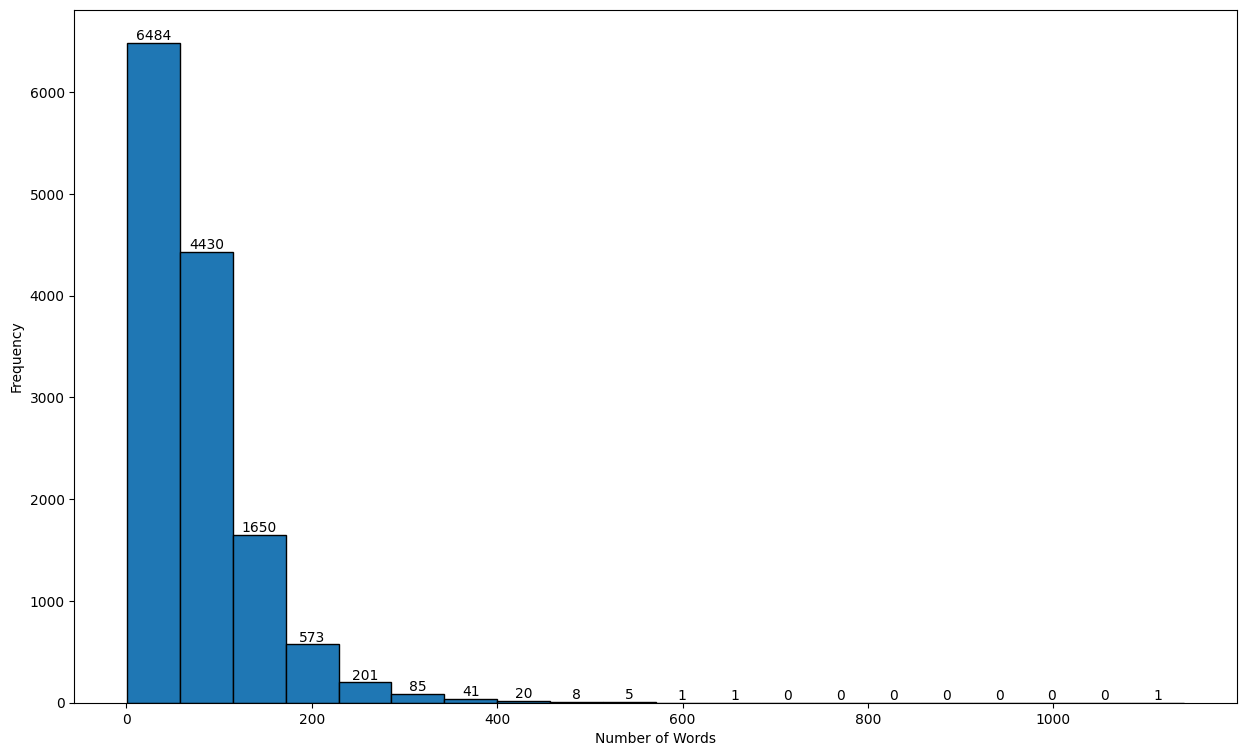}
    \caption{Distribution of Number of Words in corpus of WikiPediaRetrieval Dataset}
    \label{fig:wiki_corpus}
\end{figure}

Distribution of the number of words in the corpus and queries in WikiPediaRetrieval dataset has been shown in Figure \ref{fig:wiki_corpus} and Figure \ref{fig:wiki_queries} respectively.

\subsection{Frequently Asked Questions (FAQ)}
\label{subsec:faq}
\textbf{1) Why choose IndicTrans2 over other available translation models?}\par
\textbf{Ans: }\citeauthor{gala2023indictrans} clearly illustrate the superior performance of IndicTrans2 over other models and systems like NLLB and Google Translate for  English to Hindi tasks.  In our preliminary analysis on a subset of the BEIR datasets (Chrf scores show in Table \ref{tab:nllbvsindic} ), we also observed that IndicTrans2 outperformed alternative models, such as NLLB, in terms of translation quality.

\begin{table}[]
    \centering
    \begin{tabular}{|c|cc|}
    \hline
        \textbf{Dataset} & \textbf{IndicTrans2} & \textbf{NLLB} \\
    \hline
         Arguana & 56.30 & 37.12 \\
    \hline
         NQ & 76.13 & 60.92 \\
    \hline
    \end{tabular}
    \caption{Chrf scores between the original English text and back translated english text by the respective model for 20,000 randomly selected samples from each dataset.}
    \label{tab:nllbvsindic}
\end{table}






\end{document}